\documentclass[twocolumn,superscriptaddress,showpacs,preprintnumbers,amsmath,amssymb,10pt]{revtex4-1}

\usepackage{amsmath}
\usepackage{amssymb}
\usepackage{graphicx}
\usepackage{morefloats}
\usepackage[usenames,dvipsnames]{xcolor}
\usepackage{blkarray}
\usepackage{verbatim}
\usepackage{hyperref}
\usepackage{subfigure}

\numberwithin{equation}{section}
\hypersetup{colorlinks=true, citecolor=orange, urlcolor=cyan, linkcolor=blue}
\definecolor{ricca}{RGB}{255, 128, 0}
\newcommand{\reff}[1]{(\ref{#1})}
\newcommand{\beq}{\begin{equation}}
\newcommand{\eneq}{\end{equation}}

\newcommand{\prg}{p_\text{BiRG}}

\begin{document}
\title{Inferring monopartite projections of bipartite networks:\\ an entropy-based approach}
\author{Fabio Saracco}
\affiliation{IMT School for Advanced Studies, P.zza S. Ponziano 6, 55100 Lucca (Italy)}
\affiliation{Institute for Complex Systems (ISC-CNR) UOS Sapienza, ``Sapienza'' University of Rome, P.le A. Moro 5, 00185 Rome (Italy)}
\author{Mika J. Straka}
\affiliation{IMT School for Advanced Studies, P.zza S. Ponziano 6, 55100 Lucca (Italy)}
\author{Riccardo Di Clemente}
\affiliation{Department of Civil and Environmental Engineering, Massachusetts Institute of Technology, Massachusetts Avenue 77, MA 02139 Cambridge (USA)}
\author{Andrea Gabrielli}
\affiliation{IMT School for Advanced Studies, P.zza S. Ponziano 6, 55100 Lucca (Italy)}
\affiliation{Institute for Complex Systems (ISC-CNR) UOS Sapienza, ``Sapienza'' University of Rome, P.le A. Moro 5, 00185 Rome (Italy)}
\author{Guido Caldarelli}
\affiliation{IMT School for Advanced Studies, P.zza S. Ponziano 6, 55100 Lucca (Italy)}
\affiliation{Institute for Complex Systems (ISC-CNR) UOS Sapienza, ``Sapienza'' University of Rome, P.le A. Moro 5, 00185 Rome (Italy)}
\author{Tiziano Squartini}
\email{tiziano.squartini@imtlucca.it}
\affiliation{IMT School for Advanced Studies, P.zza S. Ponziano 6, 55100 Lucca (Italy)}
\affiliation{Institute for Complex Systems (ISC-CNR) UOS Sapienza, ``Sapienza'' University of Rome, P.le A. Moro 5, 00185 Rome (Italy)}
\date{\today}

\begin{abstract}
Bipartite networks are currently regarded as providing a major insight into the organization of many real-world systems, unveiling the mechanisms driving the interactions occurring between distinct groups of nodes. One of the most important issues encountered when modeling bipartite networks is devising a way to obtain a (monopartite) projection on the layer of interest, which preserves as much as possible the information encoded into the original bipartite structure. In the present paper we propose an algorithm to obtain statistically-validated projections of bipartite networks, according to which any two nodes sharing a statistically-significant number of neighbors are linked. Since assessing the statistical significance of nodes similarity requires a proper statistical benchmark, here we consider a set of four null models, defined within the Exponential Random Graph framework. Our algorithm outputs a matrix of link-specific p-values, from which a validated projection is straightforwardly obtainable, upon running a multiple hypothesis testing procedure. Finally, we test our method on an economic network (i.e. the countries-products World Trade Web representation) and a social network (i.e. MovieLens, collecting the users' ratings of a list of movies). In both cases non-trivial communities are detected: while projecting the World Trade Web on the countries layer reveals modules of similarly-industrialized nations, projecting it on the products layer allows communities characterized by an increasing level of complexity to be detected; in the second case, projecting MovieLens on the films layer allows clusters of movies whose affinity cannot be fully accounted for by genre similarity to be individuated.
\end{abstract}

\maketitle
\section{Introduction}\label{sec:Intro}

Many real-world systems, ranging from biological to socio-economic ones, are bipartite in nature, being defined by interactions occurring between pairs of distinct groups of nodes (be they authorships, attendances, affiliations, etc.) \cite{Caldarelli2007, Newman2010}. This is the reason why bipartite networks are ubiquitous tools, employed in many different research areas to gain insight into the mechanisms driving the organization of the aforementioned complex systems.

One of the issues encountered when modeling bipartite networks is obtaining a (monopartite) projection over the layer of interest while preserving as much as possible the information encoded into the original bipartite structure. This problem becomes particularly relevant when, e.g. a direct measurement of the relationships occurring between nodes belonging to the same layer is impractical (as gathering data on friendship within social networks \cite{Review2014}).

The simplest way of inferring the presence of otherwise unaccessible connections is linking any two nodes, belonging to the same layer, as long as they share at least one neighbor: however, this often results in a very dense network whose topological structure is almost trivial. A solution which has been proposed prescribes to retain the information on the number of common neighbors, i.e. to project a bipartite network into a \emph{weighted} monopartite network \cite{Review2014}. This prescription, however, causes the nodes with larger degree in the original bipartite network to have, in turn, larger strengths in the projection, thus masking the genuine statistical relevance of the induced connections. Moreover, such a prescription lets spurious clusters of nodes emerge (e.g. cliques induced by the presence of - even - a single node connected to all nodes on the opposite layer).

In order to face this problem, algorithms to retain only the significant weights have been proposed \cite{Review2014}. Many of them are based on a thresholding procedure, a major drawback of which lies in the arbitrariness of the chosen threshold \cite{Latapy2008,Watts1998,Derudder2005}. A more statistically-grounded algorithm prescribes to calculate the statistical significance of the projected weights according to a properly-defined null model \cite{Serrano2009}; the latter, however, encodes relatively little information on the original bipartite structure, thus being more suited to analyze natively monopartite networks. A similar-in-spirit approach aims at extracting the backbone of a weighted, monopartite projection by calculating its Minimum Spanning Tree and provides a recipe for community detection by calculating the Minimum Spanning Forest \cite{Caldarelli2012,Zaccaria2014}. However, the lack of a comparison with a benchmark makes it difficult to asses the statistical relevance of its outcome.

The approaches discussed so far represents attempts to validate a projection \emph{a posteriori}. A different class of methods, on the other hand, focuses on \emph{projecting} a statistically validated network by estimating the tendency of any two nodes belonging to the same layer to share a given portion of neighbors. All approaches define a similarity measure which either ranges between 0 and 1 \cite{Amazon.com,Bonacich1972} or follows a probability distribution on which a p-value can be computed \cite{Tumminello2011,Gualdi2016,Dianati2016}. While in the first case the application of an arbitrary threshold is still unavoidable, in the second case prescriptions rooted in traditional statistics can be applied.

In order to overcome the limitations of currently-available algorithms, we propose a general method which rests upon the very intuitive idea that any two nodes belonging to the same layer of a bipartite network should be linked in the corresponding monopartite projection if, and only if, significantly similar. To stress that our benchmark is defined by constraints which are satisfied \emph{on average}, we will refer to our method as to a \emph{grand canonical} algorithm for obtaining a statistically-validated projection of any binary, undirected, bipartite network. A \emph{microcanonical} projection method has been defined as well \cite{Guillaume2003} which, however, suffers from a number of limitations imputable to its nature of purely numerical algorithm \cite{Review2014}.
\newline
\newline
\indent The rest of the paper is organized as follows. In the Methods section, our approach is described: first, we introduce a quantity to measure the similarity of any two nodes belonging to the same layer; then, we derive the probability distribution of this quantity according to four bipartite null models, defined within the Exponential Random Graph (ERG) formalism \cite{Saracco2015}. Subsequently, for any two nodes, we quantify the statistical significance of their similarity and, upon running a multiple hypothesis test, we link them if recognized as significantly similar. In the Results section we employ our method to obtain a projection of two different data sets: the countries-products World Trade Web and the users-movies MovieLens network. Finally, in the Discussions section we comment our results.

\section{Methods}\label{sec:Methods}

A bipartite, undirected, binary network is completely defined by its biadjacency matrix, i.e. a rectangular matrix $\mathbf{M}$ whose dimensions will be indicated as $N_R\times N_C$, with $N_R$ being the number of nodes in the top layer (i.e. the number of rows of $\mathbf{M}$) and $N_C$ being the number of nodes in the bottom layer (i.e. the number of columns of $\mathbf{M}$). $\mathbf{M}$ sums up the structure of the corresponding bipartite matrix: $m_{rc}=1$ if node $r$ (belonging to the top layer) and node $c$ (belonging to the bottom layer) are linked, otherwise $m_{rc}=0$. Links connecting nodes belonging to the same layer are not allowed.

In order to obtain a (layer-specific) monopartite projection of a given bipartite network, a criterion for linking the considered pairs of nodes is needed. Schematically, our grand canonical algorithm works as follows:

\begin{itemize}
\item[{\bf A.}] choose a specific pair of nodes belonging to the layer of interest, say $r$ and $r'$, and measure their similarity;
\item[{\bf B.}] quantify the statistical significance of the measured similarity with respect to a properly-defined null model, by computing the corresponding p-value;
\item[{\bf C.}] link nodes $r$ and $r'$ if, and only if, the related p-value is statistically significant;
\item[*] repeat the steps above for every pair of nodes.
\end{itemize}

We will now describe each step of our algorithm in detail.

\subsection{Measuring nodes similarity} 

\begin{figure}[t!]
\begin{center}
\includegraphics[width=0.49\textwidth]{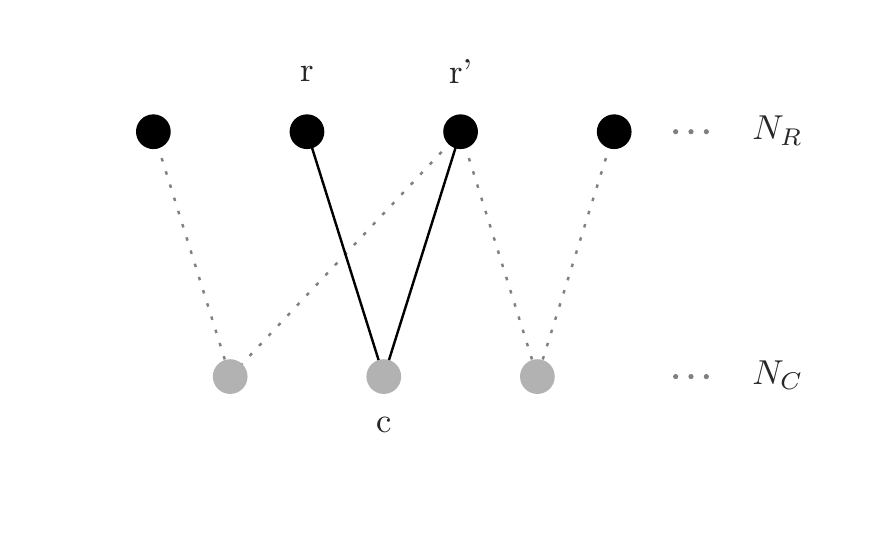}
\caption{Pictorial representation of the $V_{rr'}^c$ motif used to define our nodes similarity measure $V_{rr'}=\sum_{c=1}^{N_C} m_{rc}m_{r'c}=\sum_{c=1}^{N_C} V_{rr'}^c$.} 
\label{fig0}
\end{center}
\end{figure}

The first step of our algorithm prescribes to measure the degree of similarity of nodes $r$ and $r'$. A straightforward approach is counting the number of common neighbors $V_{rr'}$ shared by nodes $r$ and $r'$. By adopting the formalism proposed in \cite{Saracco2015}, our measure of similarity is provided by the number of bi-cliques $K_{1,2}$ \cite{Diestel2006}, also known as V-motifs \cite{Saracco2015}:

\begin{equation}
V_{rr'}=\sum_{c=1}^{N_C} m_{rc}m_{r'c}=\sum_{c=1}^{N_C} V_{rr'}^c
\end{equation}

\noindent where we have adopted the definition $V_{rr'}^c\equiv m_{rc}m_{r'c}$ for the single V-motif defined by nodes $r$ and $r'$ and node $c$ belonging to the opposite layer (see fig. \ref{fig0} for a pictorial representation). From the definition, it is apparent that $V_{rr'}^c=1$ if, and only if, both $r$ and $r'$ share the (common) neighbor $c$.

Notice that na\"ively projecting a bipartite network corresponds to considering the monopartite matrix defined as $\mathbf{V}_{rr'}^{naive}=V_{rr'}$ whose densely connected structure, described by $\mathbf{R}_{rr'}^{naive}=\Theta[V_{rr'}]$, is characterized by an almost trivial topology.

\subsection{Quantifying the statistical significance of nodes similarity} 

The second step of our algorithm prescribes to quantify the statistical significance of the similarity of our nodes $r$ and $r'$. To this aim, a benchmark is needed: a natural choice leads to adopt the ERG class of null-models \cite{Park2004, Garlaschelli2008, Squartini2011, Fronczak2012, Mastrandrea2014, Saracco2015}.

Within the ERG framework, the generic bipartite network $\mathbf{M}$ is assigned an exponential probability $P(\mathbf{M})=\frac{e^{-H(\vec{\theta},\:\vec{C}(\mathbf{M}))}}{Z(\vec{\theta})}$, whose value is determined by the vector $\vec{C}(\mathbf{M})$ of topological constraints \cite{Park2004}. In order to determine the unknown parameters $\vec{\theta}$, the likelihood-maximization recipe can be adopted: given an observed biadjacency matrix $\mathbf{M}^*$, it translates into solving the system of equations $\langle \vec{C}\rangle(\vec{\theta})=\sum_{\mathbf{M}}P(\mathbf{M})\vec{C}(\mathbf{M})=\vec{C}(\mathbf{M}^*)$ which prescribes to equate the ensemble averages $\langle \vec{C}\rangle(\vec{\theta})$ to their observed counterparts, $\vec{C}(\mathbf{M}^*)$ \cite{Garlaschelli2008}.

Two of the null models we have considered in the present paper are known as the Bipartite Random Graph (BiRG) model and the Bipartite Configuration Model (BiCM) \cite{Saracco2015}; the other ones are the two ``partial'' configuration models $\text{BiPCM}_r$ and $\text{BiPCM}_c$: the four null models are defined, respectively, by constraining the total number of links, the degrees of nodes belonging to both layers and the degrees of nodes belonging to one layer only (see Appendix for the analytical definitions).

The use of linear constraints allows us to write $P(\mathbf{M})$ in a factorized form, i.e. as the product of pair-specific probability coefficients

\begin{equation}
P(\mathbf{M})=\prod_{r=1}^{N_R}\prod_{c=1}^{N_C}p_{rc}^{m_{rc}}(1-p_{rc})^{1-m_{rc}}
\end{equation}

\noindent the numerical value of the generic coefficient $p_{rc}$ being determined by the likelihood-maximization condition (see Appendix). As an example, in the case of BiRG, $p_{rc}=p_\text{BiRG}=\frac{L}{N_R\cdot N_C},\:\forall\:r,c$ with $L$ being the total number of links in the actual bipartite network.

Since ERG models with linear constraints treat links as independent random variables, the presence of each $V_{rr'}^c$ can be regarded as the outcome of a Bernoulli trial:

\begin{eqnarray}
f_\text{Ber}(V_{rr'}^c=1)&=&p_{rc}p_{r'c},\\
f_\text{Ber}(V_{rr'}^c=0)&=&1-p_{rc}p_{r'c}.
\end{eqnarray}

It follows that, once $r$ and $r'$ are chosen, the events describing the presence of the $N_C$ single $V_{rr'}^c$ motifs are independent random experiments: this, in turn, implies that each $V_{rr'}$ is nothing else than a sum of independent Bernoulli trials, each one described by a different probability coefficient.

The distribution describing the behavior of each $V_{rr'}$ turns out to be the so-called Poisson-Binomial \cite{hoeffding1956, wang1993}. More explicitly, the probability of observing zero V-motifs between $r$ and $r'$ (or, equivalently, the probability for nodes $r$ and $r'$ of sharing zero neighbors) reads

\begin{equation}
f_{\text{PB}}(V_{rr'}=0)=\prod_{c=1}^{N_C} (1-p_{rc}p_{r'c}),
\end{equation}

\noindent the probability of observing only one V-motif reads

\begin{equation}
f_{\text{PB}}(V_{rr'}=1)=\sum_{c=1}^{N_C}\left[p_{rc}p_{r'c}\prod_{\substack{c'=1\\c'\neq c}}^{N_C}(1-p_{rc'}p_{r'c'})\right],
\end{equation}

\noindent etc. In general, the probability of observing $n$ V-motifs can be expressed as a sum of $\binom{N_C}{n}$ addenda, running on the n-tuples of considered nodes (in this particular case, the ones belonging to the bottom layer). Upon indicating with $C_n$ all possible nodes n-tuples, this probability reads

\begin{eqnarray}\label{distrib}
f_{\text{PB}}(V_{rr'}&=&n)=\nonumber\\
&=&\sum_{C_n}\left[\prod_{c_i\in C_n} p_{rc_i}p_{r'c_i}\prod_{c'_i\notin C_n }(1-p_{rc'_i}p_{r'c'_i})\right]\nonumber\\
\end{eqnarray}
(notice that the second product runs over the complement set of $C_n$).

Measuring the statistical significance of the similarity of nodes $r$ and $r'$ thus translates into calculating a p-value on the aforementioned Poisson-Binomial distribution, i.e. the probability of observing a number of V-motifs greater than, or equal to, the observed one (which will be indicated as $V_{rr'}^*$):

\begin{equation}
\text{p-value}(V_{rr'}^*)=\sum_{V_{rr'}\geq V_{rr'}^*}f_{\text{PB}}(V_{rr'}).
\end{equation}

Upon repeating such a procedure for each pair of nodes, we obtain an $N_R \times N_R$ matrix of p-values (see also Appendix). In order to speed up the numerical computation of p-values, a Python code has been made publicly available by the authors at \footnote{Python code for computing p-values under the null models discussed in the paper: \url{https://github.com/tsakim/bicm}}.

As a final remark, notice that this approach describes a one-tail statistical test, where nodes are considered as significantly similar if, and only if, the observed number of shared neighbors is ``sufficiently large''. In principle, our algorithm can be also used to carry out the reverse validation, linking any two nodes if the observed number of shared neighbors is ``sufficiently small'': this second type of validation can be performed whenever interested in highlighting the ``dissimilarity'' between nodes.

\subsection{Validating the projection}\label{FDR}

In order to understand which p-values are significant, it is necessary to adopt a statistical procedure accounting for testing multiple hypotheses at a time.

In the present paper we apply the so-called False Discovery Rate (FDR) procedure \cite{Benjamini1995}. Whenever $M$ different hypotheses, $H_1\dots H_M$, characterized by $M$ different p-values, must be tested at a time, FDR prescribes to, first, sort the $M$ p-values in increasing order, $\text{p-value}_1\leq\dots\leq \text{p-value}_M$ and, then, to identify the largest integer $\hat{i}$ satisfying the condition

\begin{equation}
\text{p-value}_{\hat{i}}\leq\dfrac{\hat{i}t}{M}
\label{threshold}
\end{equation}

\noindent with $t$ representing the usual single-test significance level (e.g. $t=0.05$ or $t=0.01$). The third step of the FDR procedure prescribes to reject all the hypotheses whose p-value is less than, or equal to, $\text{p-value}_{\hat{i}}$, i.e. $\text{p-value}_1\leq\dots\leq \text{p-value}_{\hat{i}}$. Notably, FDR allows one to control for the expected number of false ``discoveries'' (i.e. incorrectly-rejected null hypotheses), irrespectively of the independence of the hypotheses tested (our hypotheses, for example, are not independent, since each observed link affects the similarity of several pairs of nodes).

In our case, the FDR prescription translates into adopting the threshold $\hat{i}t/\binom{N_R}{2}$ which corresponds to the largest $\text{p-value}_{\hat{i}}$ satisfying the condition

\begin{equation}\label{pall}
\text{p-value}_{\hat{i}}\leq\dfrac{\hat{i}t}{\binom{N_R}{2}}
\end{equation}

\noindent (with $i$ indexing the sorted $\binom{N_R}{2}$ \text{p-value}$(V_{rr'})$ coefficients) and considering as significantly similar only those pairs of nodes $r$, $r'$ whose $\text{p-value}(V_{rr'}^*)\leq\text{p-value}_{\hat{i}}$. In other words, every couple of nodes whose corresponding p-value is validated by the FDR is joined by a binary, undirected link in our projection. In what follows, we have used a single-test significance level of $t=0.01$.

Summing up, the recipe for obtaining a statistically-validated projection of the bipartite network $\mathbf{M}$ by running the FDR criterion requires that $\mathbf{R}_{rr'}^{nm}=1$ if, and only if, $\text{p-value}(V_{rr'})\leq \text{p-value}_{\hat{i}}$, according to null model $nm$ used. Notice that the validation process naturally circumvents the problem of spurious clustering (see also Appendix).
\newline
\newline
\indent The aforementioned approaches providing an algorithm to project a validated network differ in the way the issue of comparing multiple hypotheses is dealt with. While in some approaches this step is simply missing and each test is carried out independently from the other ones \cite{Review2014,Dianati2016}, in others the Bonferroni correction is employed \cite{Tumminello2011,Gualdi2016}. Both solutions are affected by drawbacks.

The former algorithms, in fact, overestimate the number of \emph{incorrectly rejected} null hypotheses (i.e. of incorrectly validated links). A simple argument can, indeed, be provided: the probability that, by chance, at least one, out of $M$ hypotheses, is incorrectly rejected (i.e. that at least one link is incorrectly validated) is $\text{FWER}=1-(1-t)^M$ which is $\text{FWER}\simeq1$ for just $M=100$ tests conducted at the significance level of $t=0.05$.

The latter algorithms, on the other hand, adopt a criterion deemed as severely overestimating the number of \emph{incorrectly retained} null hypotheses (i.e. of incorrectly discarded links) \cite{Benjamini1995}. Indeed, if the stricter condition $\text{FWER}=0.05$ is now imposed, the threshold p-value can be derived as $\text{p-value}_{th}=t\simeq 0.05/M$ which rapidly vanishes as $M$ grows. As a consequence, very sparse (if not empty) projections are often obtained.

Naturally, deciding which test is more suited for the problem at hand depends on the importance assigned to false positive and false negatives. As a rule of thumb, the Bonferroni correction can be deemed as appropriate when {\it few} tests, out of a {\it small} number of multiple comparisons, are expected to be significant (i.e. when even a {\it single} false positive would be problematic). On the contrary, when {\it many} tests, out of a {\it large} number of multiple comparisons, are expected to be significant (as in the case of socio-economic networks), using the Bonferroni correction may, in turn, produce a too large number of false negatives, an undesired consequence of which may be the impairment of, e.g. a recommendation system.

As a final remark, we stress that an \emph{a priori} selection of the number of validated links is not necessarily compatible with the existence of a level $t$ of statistical significance ensuring that the FDR procedure still holds. As an example, let us suppose we retain only the first $k$ p-values; the FDR would then require the following inequalities to be satisfied: $\text{p-value}_k\leq kt/M$ and $\text{p-value}_{k+1}>(k+1)t/M$. This, in turn, would imply $\text{p-value}_k/k<\text{p-value}_{k+1}/(k+1)$. The aforementioned condition, however, can be easily violated by imaging a pair of subsequent p-values close enough to each other (e.g. $\text{p-value}_3=0.039$ and $\text{p-value}_4=0.040$).

\subsection{Testing the projection algorithm}

\subsubsection*{Community detection}

In order to test the performance of our method, the Louvain algorithm has been run on the validated projections of the real networks considered for the present analysis \cite{Blondel2008}. Since Louvain algorithm is known to be order-dependent \cite{Fortunato2010,staudt2016engineering}, we considered $N$ outcomes of the former, each one obtained by randomly reshuffling the order of nodes taken as input ($N$ being the network size), and chose the one providing the maximum value of the modularity. This procedure can be shown to enhance the detection of partitions characterized by a higher value of the modularity itself (a parallelized Python version of the reshuffled Louvain method is available at the public repository \footnote{\url{https://github.com/tsakim/Shuffled_Louvain}.}).

\section{Results}\label{sec:Results}

\begin{figure}[t!]
\begin{center}
\includegraphics[width=0.55\textwidth]{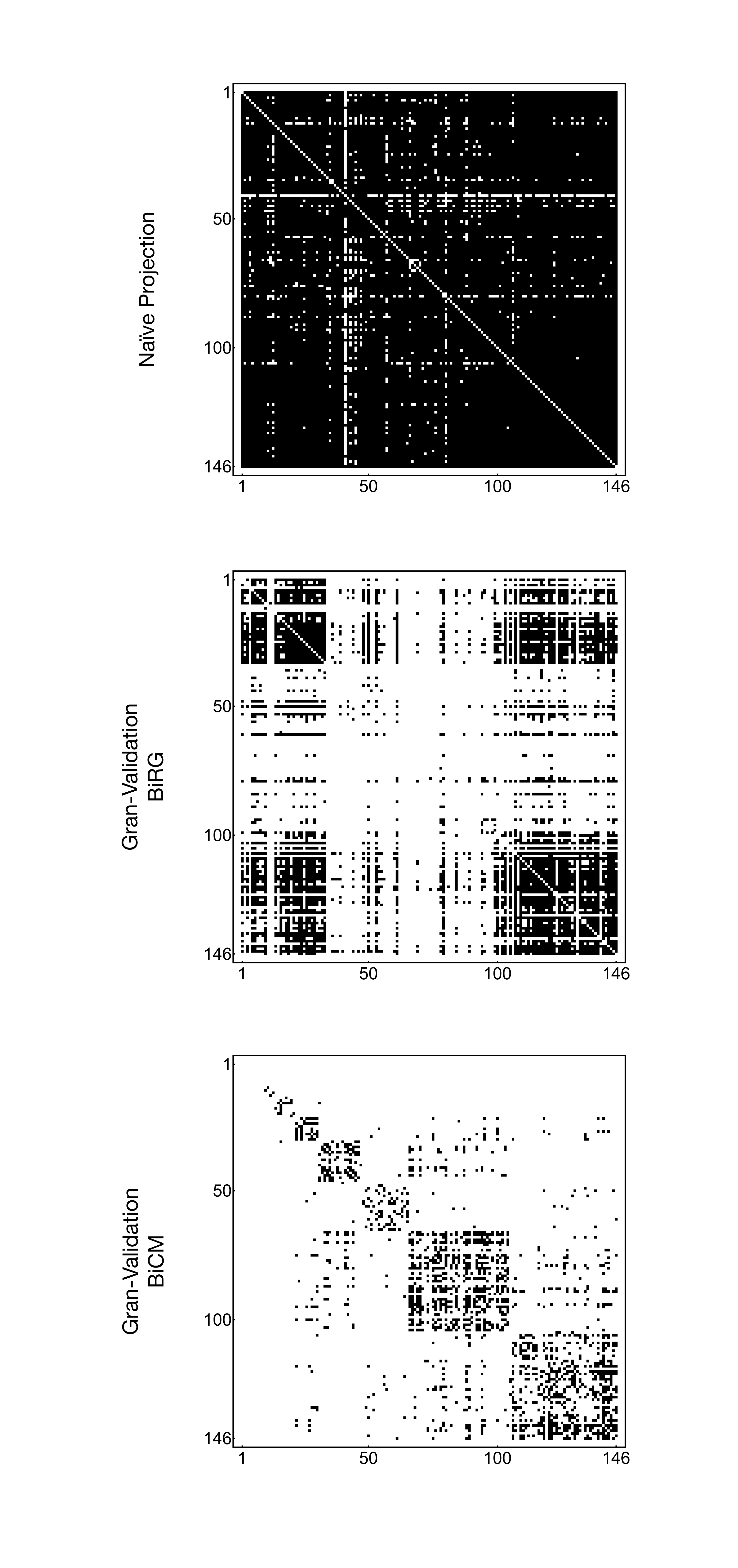}
\caption{From top to bottom, pictorial representation of the validated projections of the WTW in the year 2000 (ones are indicated as black dots, zeros as white dots): na\"ive projection $\mathbf{R}_{rr'}^{naive}$, BiRG-induced projection and BiCM-induced projection. Rows and columns of each matrix have been reordered according to the same criterion.}
\label{fig:hs2007_matrices}
\end{center}
\end{figure}

\subsection{World Trade Web}\label{ssec:hs2007}

Let us now test our validation procedure on the first data set considered for the present analysis: the World Trade Web. In the present paper we consider the COMTRADE database (using the HS 2007 code revision), spanning the years 1995-2010 \footnote{\url{http://comtrade.un.org/}}. After a data-cleaning procedure operated by BACI \cite{BACI2013} and a thresholding procedure induced by the RCA (for more details, see \cite{Tacchella2012}), we end up with a bipartite network characterized by $N_R=146$ countries and $N_C=1131$ classes of products, whose generic entry $m_{rc}=1$ indicates that country $r$ exports product $c$ above the RCA threshold.

\begin{figure*}[t!]
\begin{center}
\includegraphics[width=\textwidth]{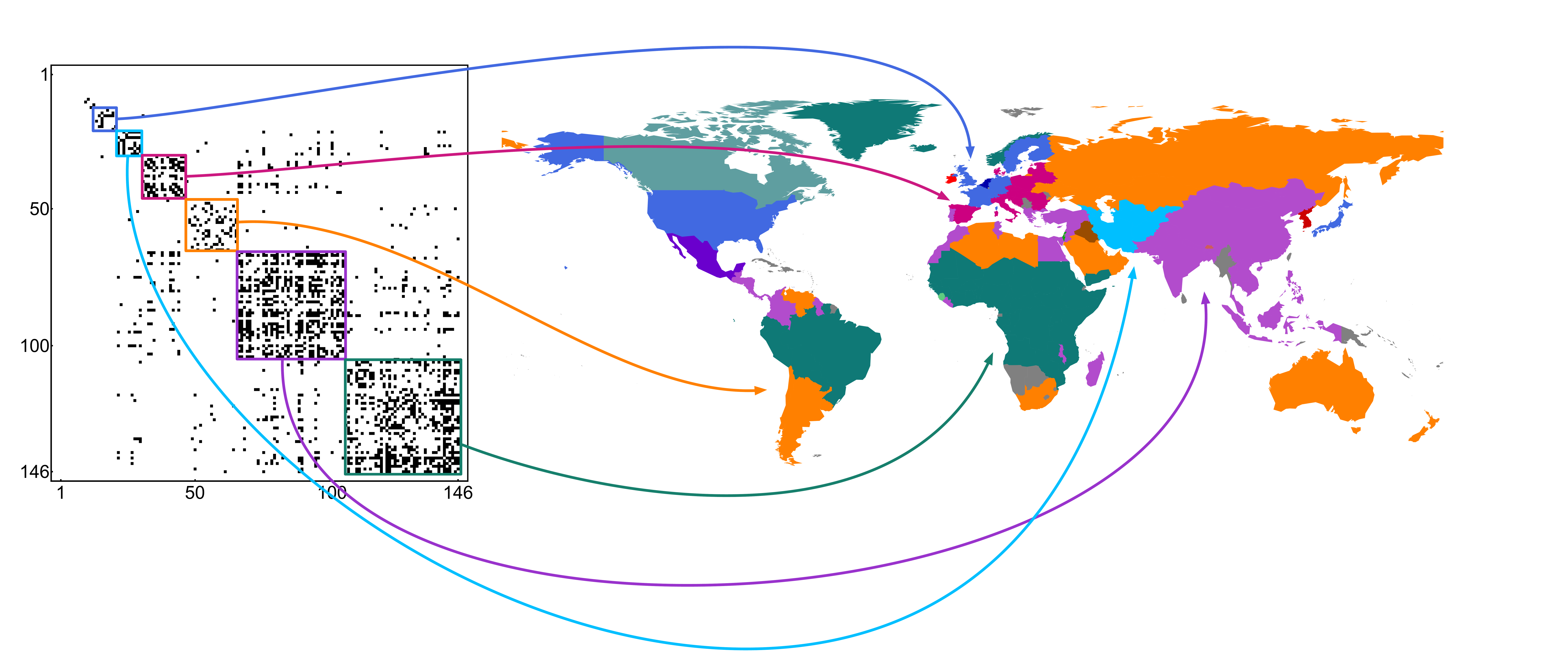}
\caption{Application of Louvain method to the BiCM-induced projection of the WTW in the year 2000. The identified communities can be interpreted as representing: \textcolor{Cerulean}{\textbullet} ``advanced'' economies (EU countries, USA and Japan, whose export basket practically includes all products); \textcolor{Orchid}{\textbullet} ``developing'' economies (centro-american countries and south-eastern countries as China, India, Asian Tigers, etc., for which the textile manufacturing represents the most important sector); countries whose export heavily rests upon raw-materials like \textcolor{orange}{\textbullet} oil (Russia, Saudi Arabia, Libya, Algeria, etc.), \textcolor{PineGreen}{\textbullet} tropical agricultural food (south-american and centro-african countries), etc. Australia, New Zealand, Chile and Argentina (whose export is based upon sea-food) happen to be detected as a community on its own.}
\label{fig:hs2007_BiCM_val001_Louvain_2000}
\end{center}
\end{figure*}

\begin{figure*}[t!]
\begin{center}
\includegraphics[width=0.49\textwidth]{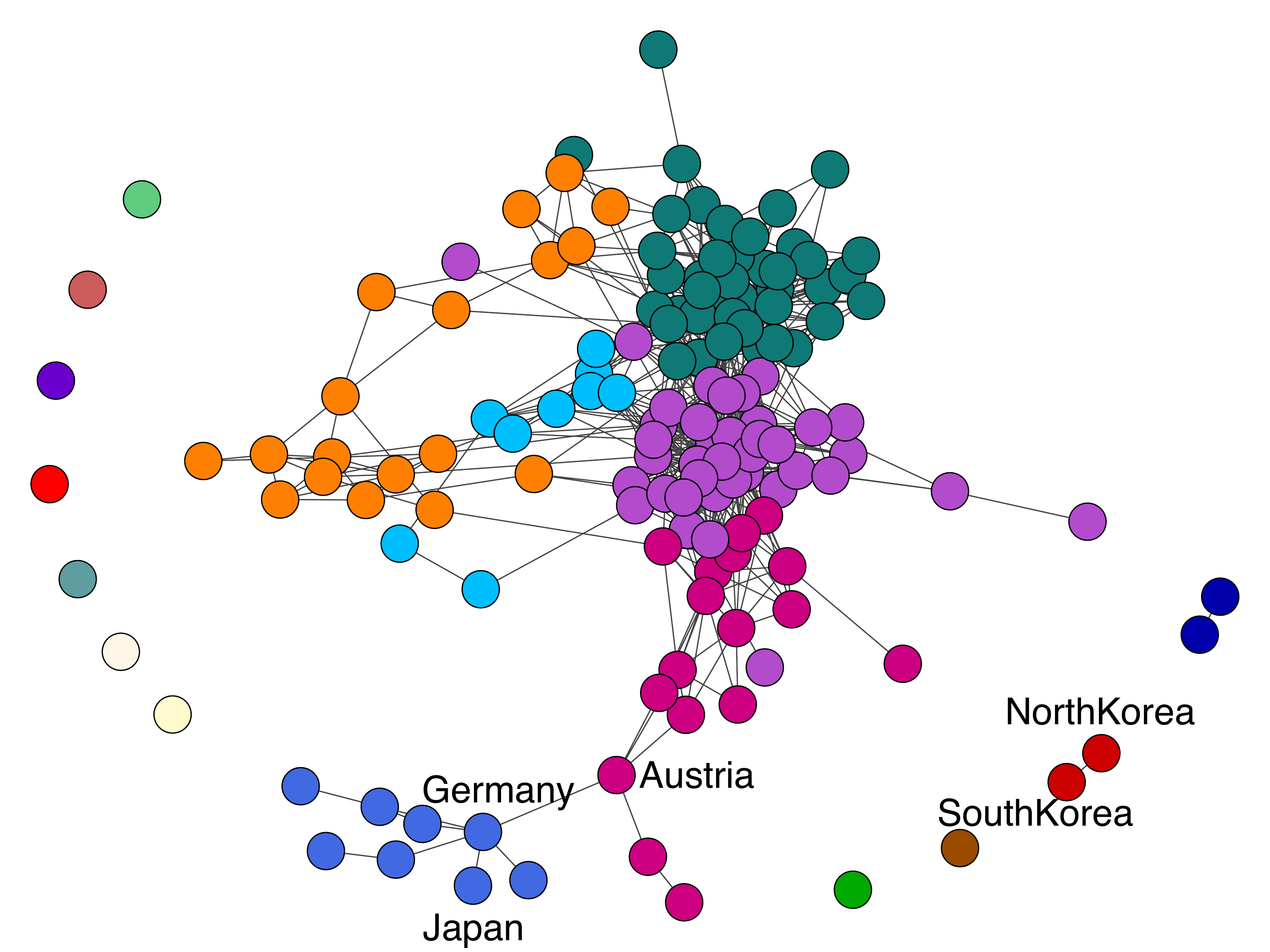}
\includegraphics[width=0.49\textwidth]{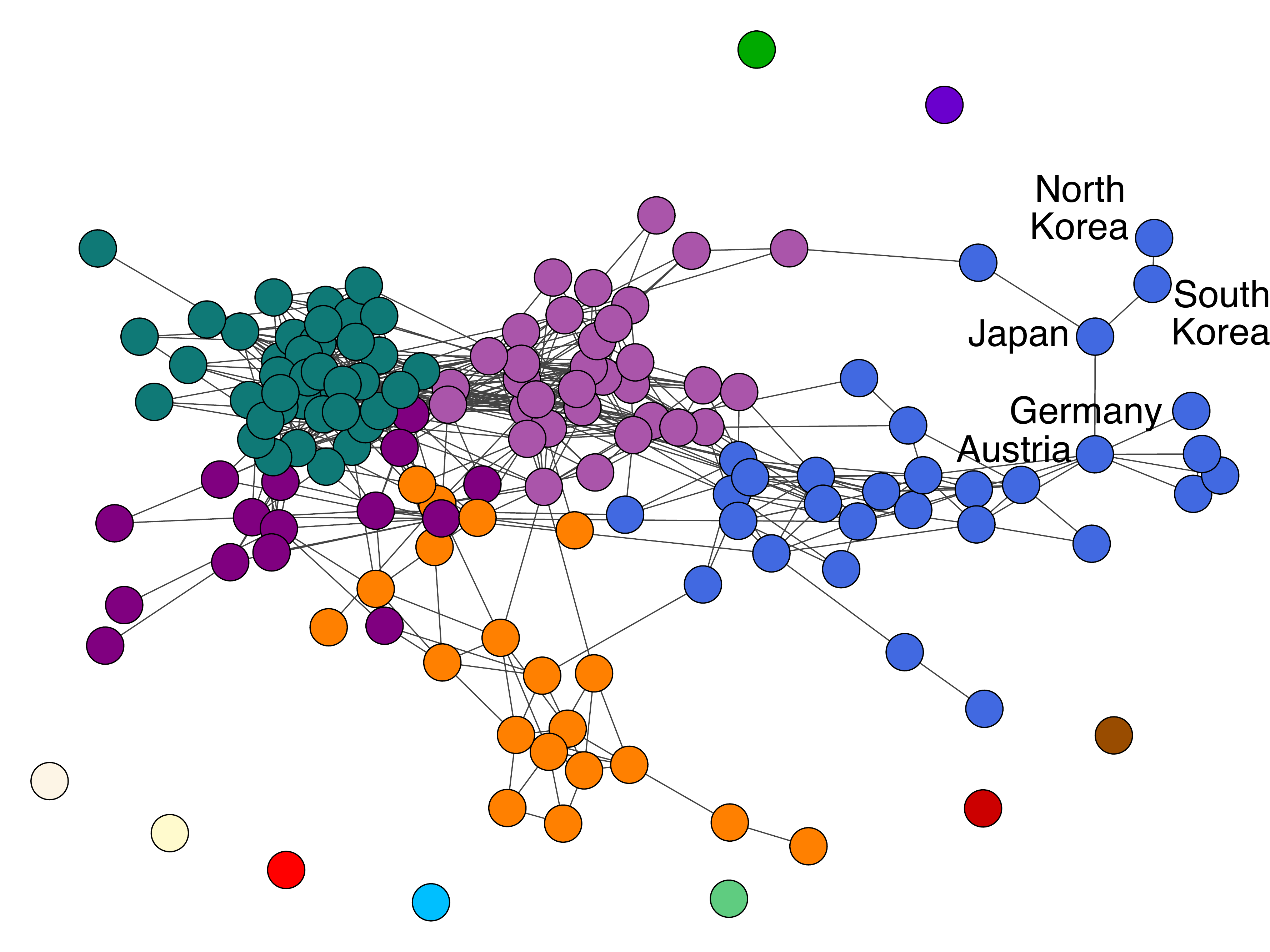}
\caption{Evolution of the topological structure of the WTW in 2000 (left panel) and 2008 (right panel). Mesoscopic patterns of self-organization emerge: the detected communities appear to be linked in a hierarchical fashion, with the ``developing'' economies seemingly constituting an intermediate layer between ``advanced'' economies and countries whose export heavily rests upon raw-materials (same colors as in fig. \ref{fig:hs2007_BiCM_val001_Louvain_2000}). Besides, the ``structural'' role played by single nodes appear: as an example, Germany is always characterized by a star-like pattern of connections which clearly indicates its prominent role in the world economy.}
\label{fig:hs2007_BiCM_val001_hc_map_2000}
\end{center}
\end{figure*}

\paragraph*{Countries layer.} Fig. \ref{fig:hs2007_matrices} shows three different projections of the WTW. The first panel shows a pictorial representation of the WTW topology in the year 2000, upon na\"ively projecting it (i.e. by joining any two nodes if at least one neighbor is shared, thus obtaining a matrix $\mathbf{R}_{rr'}^{naive}=\Theta[V_{rr'}]$). The high density of links (which oscillates between 0.93 and 0.95 throughout the period covered by the data set) causes the network to be characterized by trivial values of structural quantities (e.g. all nodes have a clustering coefficient very close to 1).

\begin{figure*}[t!]
\begin{center}
\includegraphics[width=\textwidth]{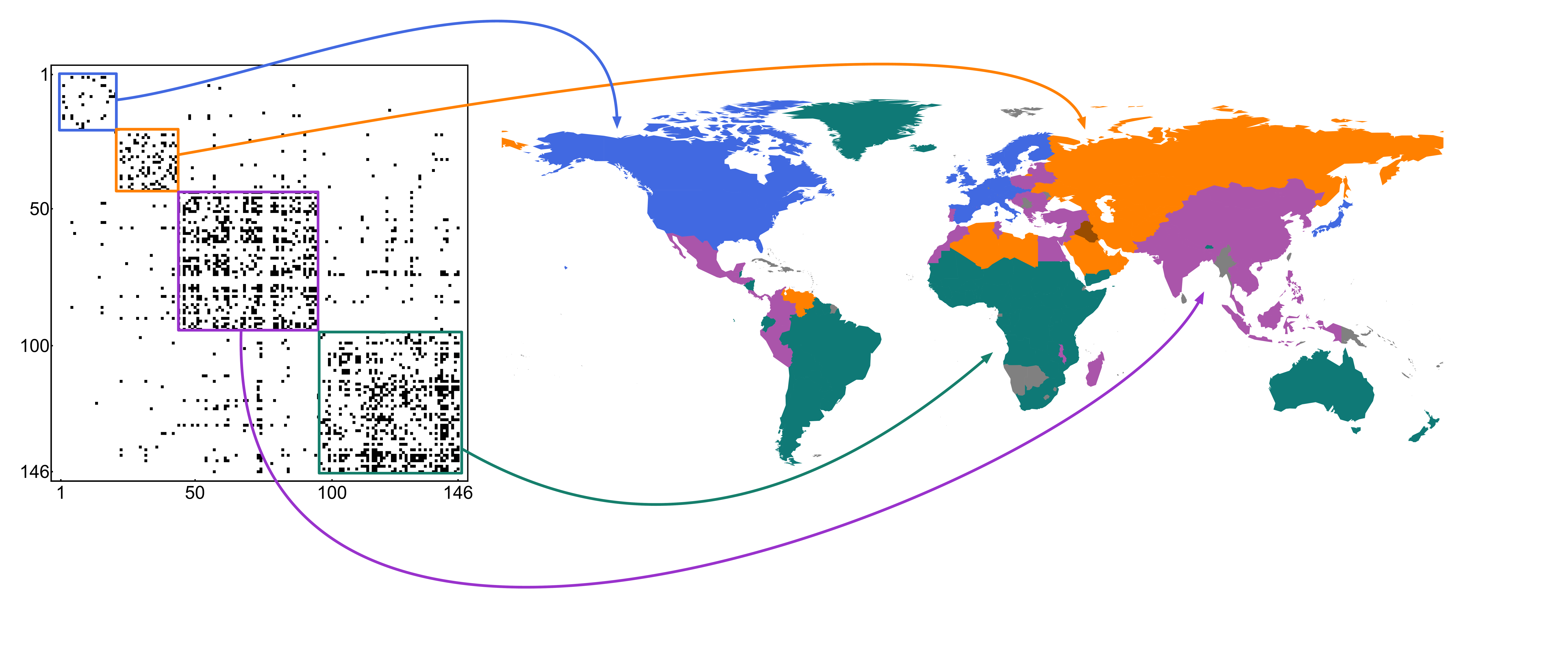}
\caption{Application of Louvain method to the $\text{BiPCM}_r$-induced projection of the WTW in the year 2000, defined by the constraints represented by the countries degrees only. Mesoscopic patterns similar to the ones revealed by the BiCM emerge, thus suggesting the $\text{BiPCM}_r$ as a computationally faster, yet equally accurate, alternative to the BiCM.}
\label{fig:hs2007_BiCM_c_val001_hc_map_2000}
\end{center}
\end{figure*}

The second panel of fig. \ref{fig:hs2007_matrices} represents the projected adjacency matrix using the BiRG as a null model. In this case, the only parameter defining our reference model is $p_\text{BiRG}=\frac{L}{N_R\cdot N_C}\simeq 0.13$. As a consequence, $p_{rc}=p_\text{BiRG}$ for every pair of nodes and formula \ref{distrib} simplifies to the Binomial

\begin{equation}
f_{\text{Bin}}(V_{rr'}=n)=\binom{N_C}{n}(p_\text{BiRG}^2)^n(1-p_\text{BiRG}^2)^{N_C-n}.
\end{equation}

The projection provided by the BiRG individuates a unique connected component of countries (notice that the two blocks at bottom-right and top-left of the panel are linked through off-diagonal connections) beside many disconnected vertices (the big white block in the center of the matrix). Interestingly, the latter represent countries whose economy heavily rests upon the presence of raw-materials (see also fig. \ref{fig:hs2007_BiCM_val001_Louvain_2000}), in turn causing each export basket to be focused around the available country-specific natural resources. As a consequence, the similarity between these countries is not significant enough to allow the corresponding links to pass the validation procedure. In other words, the BiRG-induced projection is able to distinguish between two extreme levels of economic development, thus providing a meaningful, yet too rough, filter.

On the other hand, the BiCM-induced projection (shown in the third panel of fig. \ref{fig:hs2007_matrices}), allows for a definite structure of clusters to emerge. The economic meaning of the detected diagonal blocks can be made explicit by running the Louvain algorithm on the projected network. As fig. \ref{fig:hs2007_BiCM_val001_Louvain_2000} shows, our algorithm reveals a partition into communities enclosing countries characterized by similar economic development \cite{DiClemente2014}. In particular, we recognize the ``advanced'' economies (EU countries, USA and Japan - whose export basket is practically constituted by all products \cite{Hidalgo2007,Hidalgo2009,Hausmann2010,Caldarelli2012,Tacchella2012,Cristelli2013,Tacchella2013}), the ``developing'' economies (as centro-american countries and south-eastern countries as China, India, Asian Tigers, etc., for which the textile manufacturing represents the most important sector) and countries whose export heavily rests upon raw-materials like oil (Russia, Saudi Arabia, Libya, Algeria, etc.), tropical agricultural food (south-american and centro-african countries), etc. An additional group of countries whose export is based upon sea-food is constituted by Australia, New Zealand, Chile and Argentina, which happen to be detected as a community on its own in partitions with comparable values of modularity.

Our algorithm is also able to highlight the structural changes that have affected the WTW topology across the temporal period considered for the present analysis. Fig. \ref{fig:hs2007_BiCM_val001_hc_map_2000} shows two snapshots of the WTW, referring to the years 2000 and 2008. While in 2000 EU countries were split into two different modules, with the north-european countries (as Germany, UK, France) grouped together with USA and Japan and the south-eastern european countries constituting a separate cluster, this is no longer true in 2008. Furthermore, the structural role played by single nodes is also pointed out. As an example, Austria and Japan emerge as two of the countries with highest betweenness, indicating their role of bridges respectively between western and eastern european countries and western and eastern world countries. A second example is provided by Germany, whose star-like pattern of connections clearly indicates its prominent role in the global trade.

\begin{figure*}[t!]
\begin{center}
\includegraphics[width=0.8\textwidth]{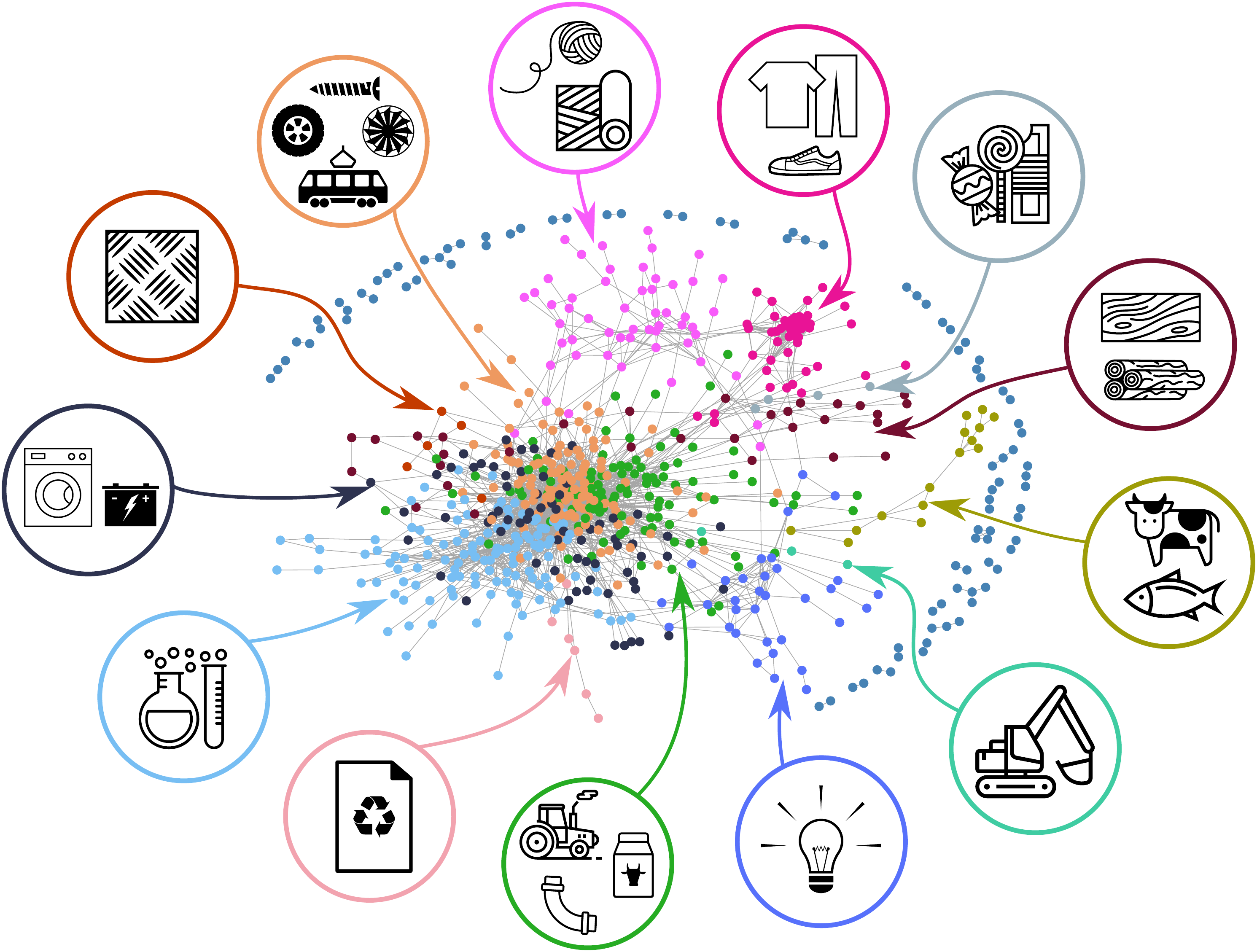}
\caption{Application of Louvain method to the $\text{BiPCM}_c$-induced projection of the WTW in the year 2000, defined by constraining the products degrees only. The identified larger communities represent: \textcolor{Rhodamine}{\textbullet} fabrics, yarn, etc.; \textcolor{Magenta}{\textbullet} clothes, shoes, etc.; \textcolor{Brown}{\textbullet} wooden products; \textcolor{GreenYellow}{\textbullet} live animals; \textcolor{Cyan}{\textbullet} basic electronics; \textcolor{Cyan}{\textbullet} chemicals; \textcolor{Black}{\textbullet} machinery; \textcolor{Orange}{\textbullet} advanced electronics (all icons are available on \url{http://thenounproject.com/} - see also [41]).}
\label{fig:hs2007_BiCM_p1_val001_hc_map_2000}
\end{center}
\end{figure*}

\begin{figure*}[t!]
\begin{center}
\includegraphics[width=0.49\textwidth]{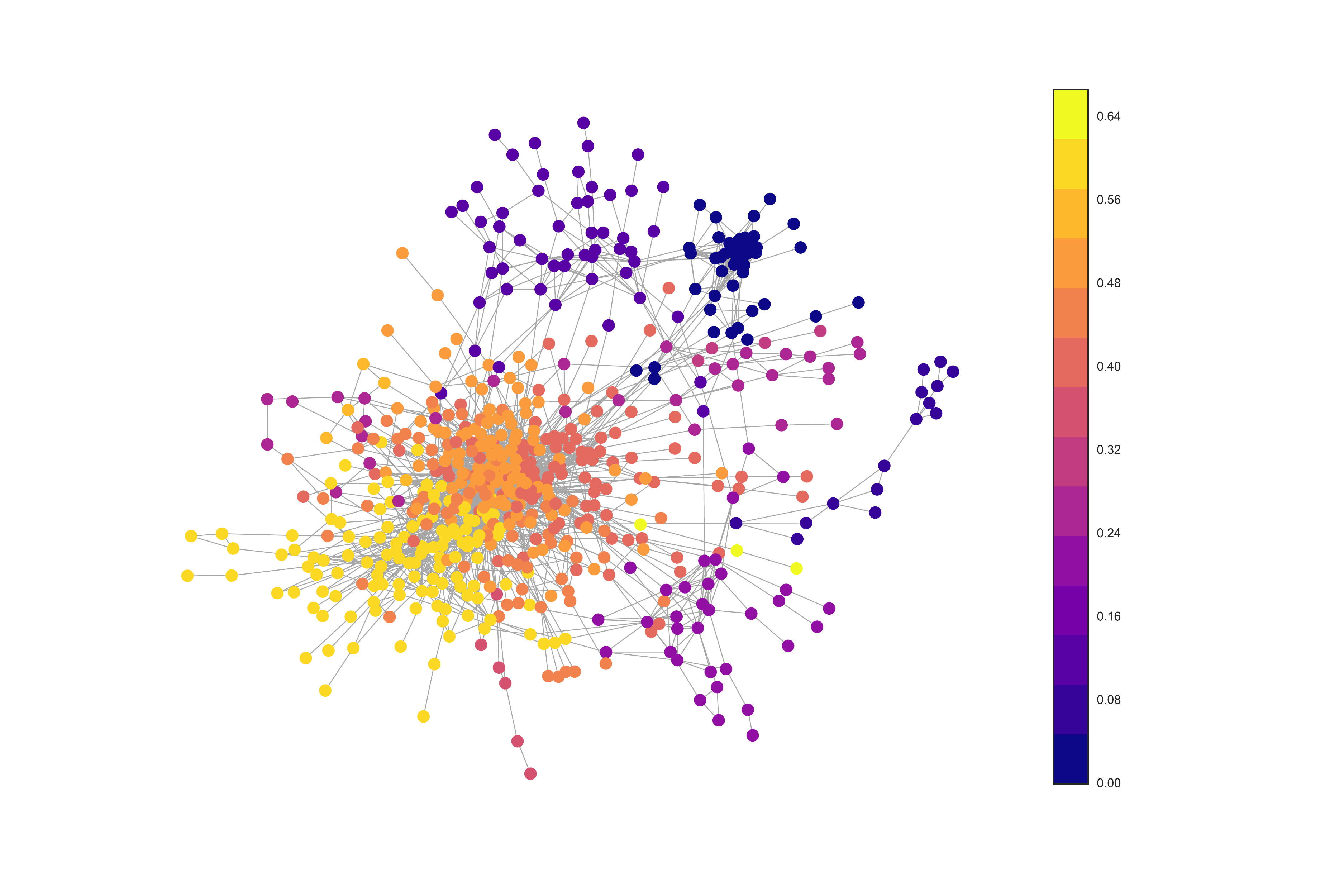}
\includegraphics[width=0.49\textwidth]{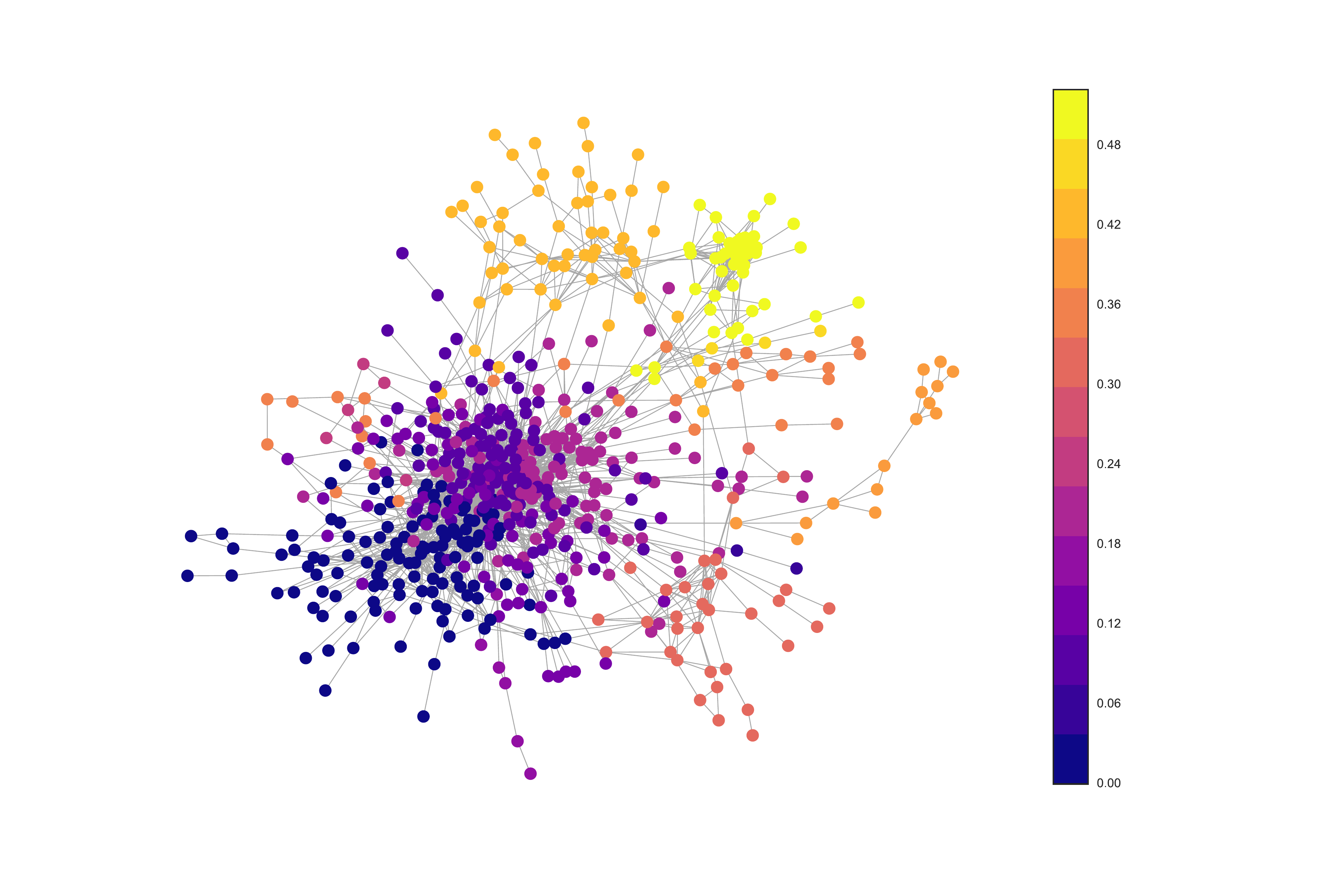}
\caption{$\text{BiPCM}_c$-induced projection of the WTW in the year 2000, with colors indicating the intensity of trade activity of ``advanced'' economies (left panel) and ``developing'' economies (right panel) over the products communities shown in fig. \ref{fig:hs2007_BiCM_p1_val001_hc_map_2000}. While the former mainly focus on high-complexity products (as chemicals, machinery, etc.), the latter mainly focus on low-complexity products (as textiles, wooden products, etc.) \cite{Tacchella2012,Cristelli2013,Tacchella2013}.}
\label{fig:hs2007_BiCM_p2_val001_hc_map_2000}
\end{center}
\end{figure*}

The block diagonal structure of the BiCM-induced adjacency matrix reflects another interesting pattern of the world economy self-organization: the detected communities appear to be linked in a hierarchical fashion, with the ``developing'' economies seemingly constituting an intermediate layer between the ``advanced'' economies and those countries whose export heavily rests upon raw-materials. Interestingly, such a mesoscopic organization persists across all years of our data set, shedding new light on the WTW evolution.

As shown in fig. \ref{fig:hs2007_BiCM_c_val001_hc_map_2000}, the results obtained by running the $\text{BiPCM}_r$ (defined by constraining only the degrees of countries) are, although less detailed, compatible with the ones obtained by running the BiCM. In this case, the $\text{BiPCM}_r$ constitutes an approximation to the BiCM, providing a computationally faster, yet equally accurate, alternative to it. On the other hand, the $\text{BiPCM}_c$ induces a projection which is close to the BiRG one, thus adding little information with respect to the latter.

\paragraph*{Products layer.} While the BiCM provides an informative benchmark to infer the presence of significant connections between countries, this is not the case when focusing on products. For this reason, we consider the $\text{BiPCM}_c$, i.e. the null model defined by constraining only the products degrees \cite{procmik}: fig. \ref{fig:hs2007_BiCM_p1_val001_hc_map_2000} shows the $\text{BiPCM}_c$-induced projection of the WTW on the layer of products \footnote{``Cow'' by Nook Fulloption; ``Fish'' by Iconic; ``Excavator'' by Kokota; ``Light bulb'' by Hopkins; ``Milk'' by Artem Kovyazin; ``Curved Pipe'' by Oliviu Stoian; ``Tractor'' by Iconic; ``Recycle'' by Agus Purwanto; ``Experiment'' by Made by Made; ``Accumulator'' by Aleksandr Vector; ``Washing Machine'' by Tomas Knopp; ``Metal'' by Leif Michelsen; ``Screw'' by Creaticca Creative Agency; ``Tram'' by Gleb Khorunzhiy; ``Turbine'' by Luigi Di Capua; ``Tire'' by Rediffusion; ``Ball Of Yarn'' by Denis Sazhin; ``Fabric'' by Oliviu Stoian; ``Shoe'' by Giuditta Valentina Gentile; ``Clothing'' by Marvdrock; ``Candies'' by Creative Mania; ``Wood Plank'' by Cono Studio Milano; ``Wood Logs'' by Alice Noir from the Noun Project. All icons are under the CC licence.}. Several communities appear, the larger ones being machinery, transportation, chemicals, electronics, textiles and live animals (a partition that seems to be stable across time).

The detected communities seem to be organized into two macro-groups: ``high-complexity'' products (on the left of the figure), including machinery, chemicals, advanced electronics, etc. and ``low-complexity'' products (on the right of the figure), including live animals, wooden products, textiles, basic electronics, etc. This macroscopic separation reflects the level of economic development of the countries trading these products. As fig. \ref{fig:hs2007_BiCM_p2_val001_hc_map_2000} clarifies, the ``advanced'' economies focus their trading activity on products characterized by high complexity, while ``developing'' economies are preferentially active on low-complexity products \cite{Tacchella2012,Cristelli2013,Tacchella2013}. A simple topological index captures this tendency: $I_{\mathcal{RC}}=\frac{\sum_{r\in\mathcal{R}}\sum_{c\in\mathcal{C}}m_{rc}}{|\mathcal{R}||\mathcal{C}|}$, i.e. the link density between groups of nodes $\mathcal{R}$ and $\mathcal{C}$, indicating one of the aforementioned communities of countries and one of the aforementioned communities of products, respectively. For example, as evident upon inspecting fig. \ref{fig:hs2007_BiCM_p2_val001_hc_map_2000}, ``advanced'' economies (left panel) and ``developing'' economies (right panel) are active on different clusters of products: while the trading activity of the former is mainly constituted by, e.g. chemicals and machinery, the latter mainly trade textiles, wooden products, etc.

\subsection{MovieLens}\label{ssec:MovieLens}

Let us now consider the second data set: MovieLens 100k. MovieLens is a project by GroupLens \cite{GroupLens}, a research lab at the University of Minnesota. Data (collected from September 19th, 1997 through April 22nd, 1998) consist of $10^5$ ratings - from 1 to 5 - given by $N_C=943$ users to $N_R=1559$ different movies \footnote{\url{http://movielens.org/}.}; information about the movies (date of release and genre) and about the users (age, gender, occupation and US zip code) is also provided. We binarize the dataset by setting $m_{rc}=1$ if user $c$ rated movie $r$ at least 3, providing a favorable recension.

\begin{figure}[t!]
\begin{center}
\includegraphics[width=0.45\textwidth]{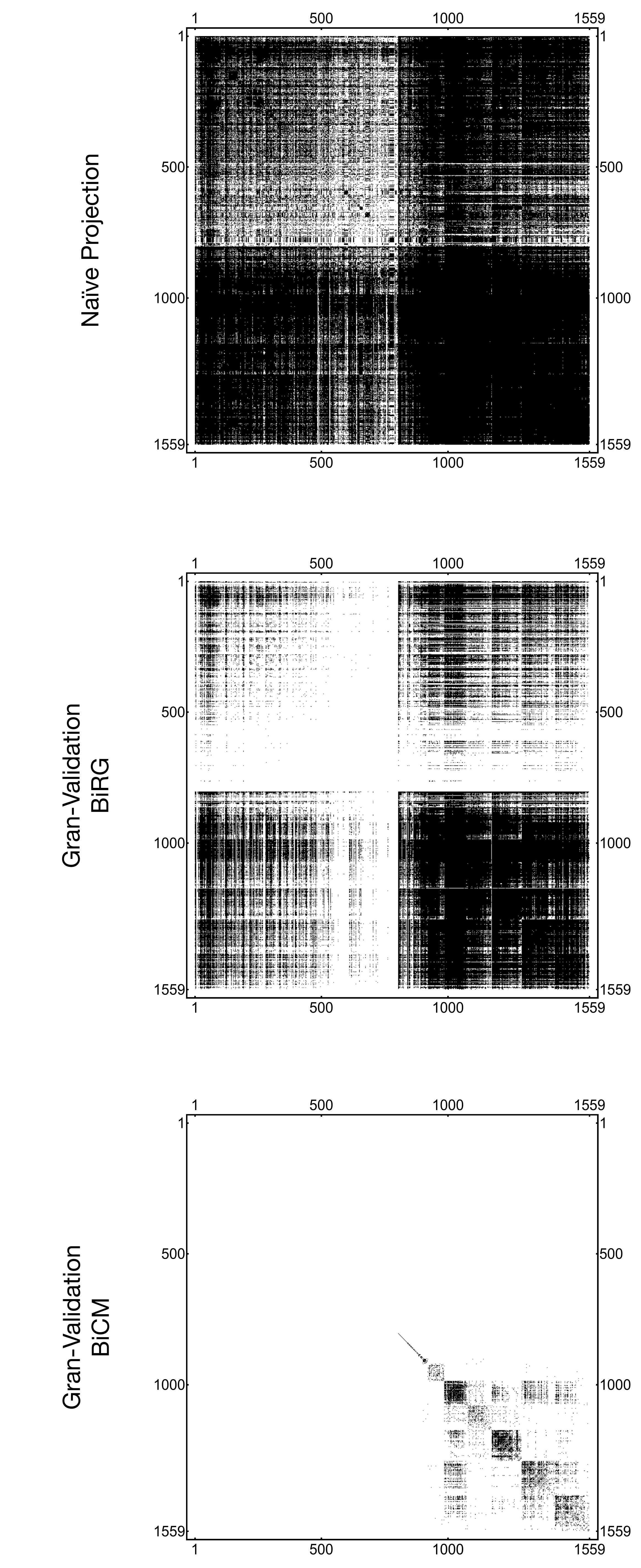}
\caption{From top to bottom, pictorial representation of the validated projections of MovieLens (ones are indicated as black dots, zeros as white dots): na\"ive projection $\mathbf{R}_{rr'}^{naive}$, BiRG-induced projection and BiCM-induced projection. Rows and columns of each matrix have been reordered according to the same criterion.}
\label{fig:MovieLens_connectance}
\end{center}
\end{figure}

In what follows we will be interested into projecting this network on the layer of movies. Fig. \ref{fig:MovieLens_connectance} shows the three projections already discussed for the WTW. As for the latter, $\mathbf{R}_{rr'}^{naive}=\Theta[V_{rr'}]$ is still a very dense network, whose connectance amounts to $0.58$. Similarly, the projection induced by the BiRG provides a rather rough filter, producing a unique large connected component, to which only the most popular movies (i.e. the ones with a large degree in the original bipartite network) belong.

\begin{figure*}[t!]
\begin{center}
\includegraphics[width=\textwidth]{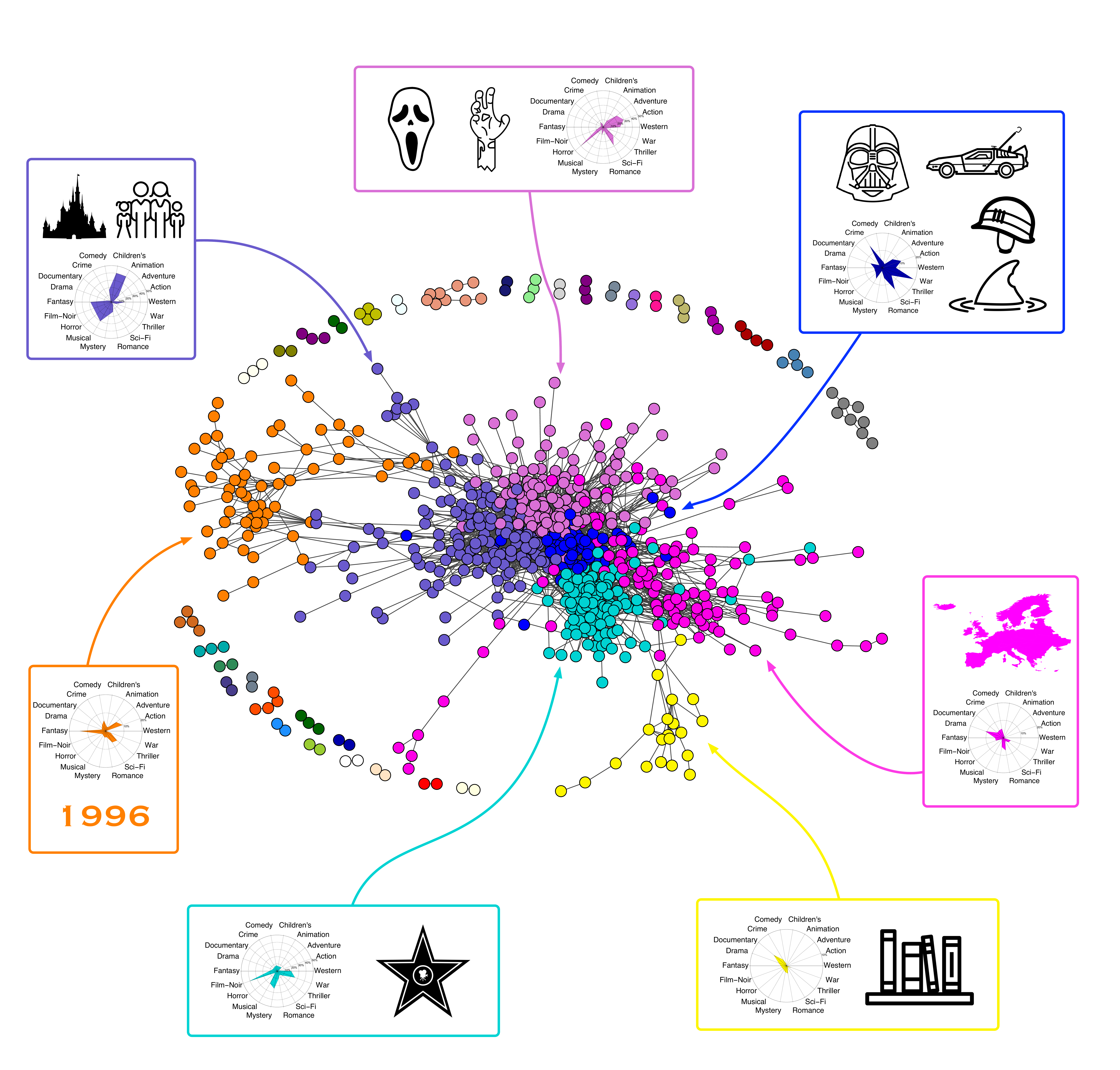}
\caption{Result of the application of Louvain method to the BiCM-induced projection of the MovieLens data set. Since some genres are quite generic, our clusters are often better described by ``combinations'' of genres (readable on the radar-plots beside them) capturing users' tastes to a larger extent: \textcolor{Orange}{\textbullet} movies released in 1996; \textcolor{CadetBlue}{\textbullet} ``family'' movies; \textcolor{Orchid}{\textbullet} movies with marked horror traits; \textcolor{Blue}{\textbullet} ``cult mass'' movies; \textcolor{Magenta}{\textbullet} independent and foreign movies; \textcolor{Yellow}{\textbullet} movies inspired to books or theatrical plays; \textcolor{Cyan}{\textbullet} ``classic'' Hollywood movies (all icons are available on \url{http://thenounproject.com/} - see also [44]).}
\label{fig:MovieLens_Lou_genres}
\end{center}
\end{figure*}

While both the na\"ive and the BiRG-induced projections only allow for a trivially-partitioned structure to be observed, this is not the case for the BiCM. By running the Louvain algorithm, we found a very composite community structure (characterized by a modularity of $Q\simeq0.58$), pictorially represented by the diagonal blocks visible in the third panel of fig. \ref{fig:MovieLens_connectance}. The BiCM further refines the results found by the BiRG, allowing for the internal structure of the blocks to emerge: in our dicussion, we will focus on the bottom-right block, which shows the richest internal organization.

Fig. \ref{fig:MovieLens_Lou_genres} shows the detected communities within the aforementioned block, beside the genres (provided together with the data) \footnote{``DeLorean'' by Aaron Humphreys; ``Darth Vader'' by Jake Dunham; ``Castle'' by Olly Banham; ``Movie Star'' by Nikita Kozin; ``Books on a Shelf'' by Lucas Glenn; ``Shark'' by Randomhero; ``Mask'' by Gorka Cestao; ``Zombie Hand'' by Valery; ``Army Helmet'' by Henry Ryder; ``Family'' by abeldb, from the Noun Project. All icons are under the CC licence.}: Action, Adventure, Animation, Children's, Comedy, Crime, Documentary, Drama, Fantasy, Horror, Musical, Mystery, Noir, Romance, Sci-Fi, Thriller, War, Western \footnote{Every movie is assigned an array of 17 entries, representing the aforementioned genres. Each entry can be either zero or one, depending if that movie is considered as belonging to that genre or not (the number of ones in the vector can vary from 1 to a maximum of 6, if the selected film falls under several genres).}. Since some genres are quite generic and, thus, appropriate for several movies (e.g. Adventure, Comedy and Drama), our clusters are often better described by ``combinations'' of genres, capturing the users' tastes to a larger extent: the detected communities, in fact, partition the set of movies quite sharply, once appropriate combinations of genres are considered.

As an example, the orange block on the left side of our matrix is composed by movies released in 1996 (i.e. the year before the survey). Remarkably, our projection algorithm is able to capture the peculiar ``similarity'' of these movies, not trivially related to the genres to which they are ascribed to (that are quite heterogeneous: Action, Comedy, Fantasy, Thriller, Sci-Fi) but to the curiosity of users towards the yearly new releases.

Proceeding clockwise, the violet block next to the orange one is composed by movies classified as Animation, Children's, Fantasy and Musical (e.g. ``Mrs. Doubtfire'', ``The Addams Family'', ``Free Willy'', ``Cinderella'', ``Snow White''). In other words, we are detecting the so-called ``family movies'', a more comprehensive definition accounting for all elements described by the single genres above.

The next purple block is composed by genres Action, Adventure, Horror, Sci-Fi and Thriller: examples are provided by ``Stargate'', ``Judge Dredd'', ``Dracula'', ``The Evil Dead''. This community encloses movies with marked horror traits, including titles far from ``mainstream'' movies. This is the main difference with respect to the following blue block: although characterized by similar genres (but with Crime replacing Horror and Thriller) movies belonging to it are more popular: ``cult mass'' movies, in fact, can be found here. Examples are provided by ``Braveheart'', ``Blade Runner'' and sagas as ``Star Wars'' and ``Indiana Jones''.

The following two blocks represent niche movies for US users. The module in magenta is, in fact, composed by foreign movies (mostly European - French, German, Italian, English - which usually combine elements from Comedy and elements from Drama), as well as US independent films (as titles by Jim Jarmush); the yellow module, on the other hand, is composed by movies inspired by books or theatrical plays and documentaries.

The last, cyan block is composed by movies which are considered as ``classic'' Hollywood movies (because of the presence of either iconic actors or master directors): examples are provided by ``Casablanca'', ``Ben Hur'', ``Taxi Driver'', ``Vertigo'' (and all movies directed by Hitchcock), ``Manhattan'', ``Annie Hall''.

As in the WTW case, running the $\text{BiPCM}_r$ (defined by constraining only the degrees of movies) leads us to obtain a coarse-grained (i.e. still informative, although less detailed) version of the aforementioned results. Only three macro-groups of movies are, in fact, detected: ``authorial'' movies (as ``classic'' Hollywood movies, Hitchcock's, Kubrick's, Spielberg's movies), recent mainstream ``blockbusters'' (as ``Star Trek'', ``Star Wars'', ``Indiana Jones'', ``Batman'' sagas) and independent/niche movies (as Spike Lee's and European movies).
\newline
\newline
\indent As a final remark, we point out that projecting on the users layer with the BiCM indeed allows several communities to be detected. However, interestingly enough, none of them seems to be accurately described by the provided indicators (age, gender, occupation and US zip code), thus suggesting that users tastes are correlated with hidden (sociometric) variables yet to be individuated.

\section{Discussion}\label{sec:Conclusion}

Projecting a bipartite network on one of its layers poses a number of problems for which several solutions have been proposed so far \cite{Amazon.com,Hidalgo2007,Tumminello2011,Caldarelli2012,Zaccaria2014,Review2014,Gualdi2016,Dianati2016}, differing from each other in the way the information encoded into the bipartite structure is dealt with.

The present paper proposes an algorithm that prescribes to, first, quantify the similarity of any two nodes belonging to the layer of interest and, then, link them if, and only if, this value is found to be statistically significant. The links constituting the monopartite projection are, thus, inferred from the co-occurrences observed in the original bipartite network, by comparing them with a proper statistical benchmark.

Since the null models considered for the present analysis retain a different amount of information, the induced projections are characterized by a different level of detail. In particular, the BiRG represents a very rough filter which employs the same probability distribution to validate the similarity between any two nodes, thereby preferentially connecting nodes with large degree than nodes with small degree. By enforcing stronger constraints (increasing the amount of retained information), stricter benchmark models are obtained.

The two partial configuration models constitute the simplest examples of benchmarks retaining also the information on the nodes degrees. However, it should be noticed that the two BiPCMs perform quite differently. In fact, the BiPCM constraining the degrees of the \emph{opposite} layer we are interested in finding a projection of, provides an homogeneous benchmark as well (i.e. the same Poisson-Binomial distribution for all pairs of nodes - see also Appendix), whence the expected little difference with respect to the BiRG performance; on the other hand, the BiPCM constraining the degrees of nodes belonging to the \emph{same} layer we are interested in finding a projection of, provides a performance which is halfway between the BiRG one and the BiCM one. The reason lies in the fact that a (Binomial) pair-specific distribution is now induced by the constraints, i.e. a benchmark properly taking into account the heterogeneity of the considered nodes. As shown in the Results section, this often allows one to obtain an accurate enough approximation to the BiCM, i.e. the null model constraining the whole degree sequence.

As also suggested in \cite{Review2014}, the use of a benchmark which ensures that the heterogeneity of all nodes is correctly accounted for is recommended: in other words, any suitable null model for projecting a network on a given layer should (at least) constrain the degree sequence of the same layer. The use of partial null models is allowed in case of constraints redundancy, e.g. when node degrees are well described by their mean (as indicated by the coefficient of variation, for example - see also Appendix): in cases like these, specifying the whole degree sequence is actually unnecessary.

As a final remark, we explicitly notice that implementing the BiCM can be computationally demanding: this is the reason why several approximations to the Poisson-Binomial distribution have been proposed so far. However, the applicability of each approximation is limited and, whenever employed to find the projection of a real, bipartite network, they may even fail to a large extent (see Appendix). With the aim of speeding up the numerical computation of the p-values induced by any of the null models discussed in the paper - while retaining the \emph{exact} expression of the corresponding distributions - a Python code has been made publicly available by the authors at [25].

Remarkably, our method can be extended in a variety of directions, e.g. to analyze directed and weighted bipartite networks, and generalized to account for co-occurrences between more than two nodes, a study that constitutes the subject of future work.

\appendix

\section*{Appendix}

\section{the Poisson-Binomial distribution}\label{sec:PBD}

The Poisson-Binomial distribution is the generalization of the usual Binomial distribution when the single Bernoulli trials are characterized by different probabilities.

More formally, let us consider $N$ Bernoulli trials, each one described by a random variable $x_i,\:i=1\dots N$, characterized by a probability of success equal to $f_{\text{Ber}}(x_i=1)=p_i$: the random variable described by the Poisson-Binomial distribution is the sum $X=\sum_i x_i$. Notice that if all $p_i$ are equal the Poisson-Binomial distribution reduces to the usual Binomial distribution.

Since every event is supposed to be independent, the expectation value of $X$ is simply

\begin{equation}\label{eq:mu}
\langle X\rangle=\sum_{i=1}^N p_i=\mu
\end{equation}
and higher-order moments read

\begin{eqnarray}\label{eq:sigmagamma}
\sigma^2 &=& \sum_{i=1}^N p_i(1-p_i),\nonumber\\
\gamma &=& \sigma^{-3}\sum_{i=1}^N p_i(1-p_i)(1-2p_i),
\end{eqnarray}

\noindent where $\sigma^2$ is the variance and $\gamma$ is the skewness.

In the problem at hand, we are interested in calculating the probability of observing a number of V-motifs larger than the measured one, i.e. the p-value corresponding to the observed occurrence of V-motifs. This translates into requiring the knowledge of the Survival Distribution Function (SDF) for the Poisson-Binomial distribution, i.e. $S_{\text{PB}}(X^*)=\sum_{X\geq X^*}f_{\text{PB}}(X)$. Reference \cite{Hong2013} proposes a fast and precise algorithm to compute the Poisson-Binomial distribution, which is based on the characteristic function of the Poisson-Binomial distribution. Let us will briefly review the main steps of the algorithm in \cite{Hong2013}. If we have observed exactly $X^*$ successes, then

\begin{eqnarray}
\text{p-value}(X^*)&=&S_{\text{PB}}(X^*)=\sum_{X\geq X^*}f_{\text{PB}}(X)=\nonumber\\
&=&\sum_{X=X^*}^N\sum_{C_X}\left[\prod_{c_i\in C_X}p_{c_i}\prod_{c_j\notin C_X}(1-p_{c_j})\right]\nonumber\\
\end{eqnarray}
where summing over $C_X$ means summing over each set of X-tuples of integers.

The problem lies in calculating $C_X$. In order to avoid to explicitly consider all the possible ways of extracting a number of $X$ integers from a given set, let us consider the Inverse Discrete Fourier Transform of $f_{\text{PB}}(X)$, i.e. 

\begin{equation}
\chi_l=\sum_{X=0}^Nf_{\text{PB}}(X)e^{i\omega Xl},
\end{equation}

\noindent \color{black}with $\omega=\frac{2\pi}{N+1}$. By comparing $\chi_l$ with the Inverse Discrete Fourier Transform of the characteristic function of $f_{\text{PB}}$, it is possible to prove (see \cite{Hong2013} for more details) that the real and the imaginary part of $\chi_l$ can be easily computed in terms of the coefficients $\{p_i\}_{i=1}^N$, which are the data of our problem: more specifically, if $z_i(l)=1-p_i+p_i \cos(\omega l)+{\mathbf i}(p_i \sin(\omega l))$, it is possible to prove that

\begin{eqnarray}
\text{Re}(\chi_l) &=& e^{\sum_{j=1}^N \log|z_j(l)|}\cos\left(\sum_{i=1}^N \text{arg}[z_i(l)]\right)\:\:\:\\
\text{Im}(\chi_l) &=& e^{\sum_{j=1}^N \log|z_j(l)|}\sin\left(\sum_{i=1}^N \text{arg}[z_i(l)]\right)\:\:\:
\end{eqnarray}

\noindent where $\text{arg}[z_i(l)]$ is the principal value of the argument of $z_i(l)$ and $|z_i(l)|$ represents its modulus. Once all terms of the Discrete Fourier Transform of $\chi_l$ (i.e. the coefficients $f_{\text{PB}}(X)$) have been derived, $S_{\text{PB}}(X)$ can be easily calculated. To the best of our knowledge, the approach proposed by \cite{Hong2013} does not suffer from the numerical instabilities which, instead, affect \cite{Chen1998}.

\section{approximations of the Poisson-Binomial distribution}\label{sec:PBDAs}

\paragraph*{Binomial approximation.} Whenever the probability coefficients of the $N$ Bernoulli trials coincide (i.e. $p_{i}=p$ as in the case of the BiRG - see later), each pair-specific Poisson-Binomial distribution reduces to the usual Binomial distribution. Notice that, in this case, all distributions coincide since the parameter is the same.

However, the Binomial approximation may also be employed whenever the distribution of the probabilities of the single Bernoulli trials is not too broad (i.e. $\sigma/\mu<0.5$): in this case, all events can be assigned the same probability coefficient $\overline{p}$, coinciding with their average $\overline{p}=\dfrac{\mu}{N}$. In this case,

\begin{equation}
S_{\text{PB}}(X)=S_{\text{Bin}}\left(X; \overline{p}, N\right).
\end{equation}

\noindent where $S_{\text{Bin}}(X; \overline{p}, N)$ is the SDF for the random variable $X$ following a binomial distribution with parameter $\overline{p}$.
\newline
\newline
\indent Whenever the aforementioned set of probability coefficients can be partitioned into homogeneous subsets (i.e. subsets of coefficients assuming the same value), the Poisson-Binomial distribution can be computed as the distribution of a sum of Binomial random variables \cite{Gualdi2016}. Such an algorithm is particularly useful when the number of subsets is not too large, a condition which translates into requiring that the heterogeneity of the degree sequences is not too high. However, when considering real networks this is often not the case and different approximations may be more appropriate.

\paragraph*{Poissonian approximation.} According to the error provided by Le Cam's theorem (stating that $\sum_{X=0}^N|f_{\text{PB}}(X)-f_{\text{Poiss}}(X)|<2\sum_{i=1}^Np_i^2$), Poisson approximation is known to work satisfactorily whenever the expected number of successes is small. In this case

\begin{equation}
S_\text{PB}(X)\simeq S_\text{Poiss}(X)
\end{equation}

\noindent where the considered Poisson distribution is defined by the parameter $\mu$ \cite{Hong2013}.

\paragraph*{Gaussian approximation.} The Gaussian approximation consists in considering

\begin{equation}
S_\text{PB}(X)\simeq S_\text{Gauss}\left(\frac{X+0.5-\mu}{\sigma}\right),
\end{equation}

\noindent where $\mu$ and $\sigma$ have been defined in \reff{eq:mu} and \reff{eq:sigmagamma}. The value 0.5 represents the continuity correction \cite{Hong2013}. Since the Gaussian approximation is based upon the Central Limit Theorem, it works in a complementary regime with respect to the Poissonian approximation: more precisely, when the expected number of successes is large.

\paragraph*{Skewness-corrected Gaussian approximation.} Based on the results of \cite{Deheuvels1989,Volkova1996}, the Gaussian approximation of the Poisson-Binomial distribution can be further refined by introducing a correction based on the value of the skewness. Upon defining

\begin{equation}\label{eq:DPR_G_function}
G(x)\equiv S_\text{Gauss}(x)-\gamma\left(\dfrac{1-x^2}{6}\right)f_\text{Gauss}(x),
\end{equation}

\noindent where $f_\text{Gauss}(x)$ is the probability density function of the standard normal distribution and $\gamma$ is defined by \reff{eq:sigmagamma}, then

\begin{equation}
S_{\text{PB}}(X)\simeq G\left(\frac{X+0.5-\mu}{\sigma}\right).
\end{equation}

The refinement described by formula \reff{eq:DPR_G_function} provides better results than the Gaussian approximation when the number of events is small.
\newline
\newline
\indent However, upon comparing the WTW projection (at the level $t=0.01$, for the year 2000) obtained by running the skewness-corrected Gaussian approximation with the projection based on the full Poisson-Binomial distribution, we found that $\simeq20\%$ of the statistically-significant links are lost in the Gaussian-based validated projection. The limitations of the Gaussian approximations are discussed in further detail in \cite{Volkova1996,Mikha1990}.

\section{null models}

\subsection{Bipartite Random Graph model}

The BiRG (Bipartite Random Graph) model is the Random Graph Model solved for bipartite networks. This model is defined by a probability coefficient for any two nodes, belonging to different layers, to connect which is equal for all pairs of nodes. More specifically, $\prg=\frac{L}{N_R\cdot N_C}$, where $L=\sum_{r=1}^{N_R}\sum_{c=1}^{N_C} m_{rc}$ is the observed number of links and $N_R$ and $N_C$ indicate, respectively, the number of rows and columns of our network. Since all probability coefficients are equal, the probability of a single V-motif (defined by the pair of nodes $r$ and $r'$ belonging to the same layer and node $c$ belonging to the second one) reads

\begin{equation}
p(V^c_{rr'})=\prg^2,
\end{equation}

Thus, the probability distribution of the number of V-motifs shared by nodes $r$ and $r'$ is simply a Binomial distribution defined by a probability coefficient equal to $\prg^2$: 

\begin{equation}
f_\text{Bin}(V_{rr'}=n)=\binom{N_C}{n}(\prg^2)^{n}(1-\prg^2)^{N_C-n}.
\end{equation}

\subsection{Bipartite Configuration Model}

The BiCM (Bipartite Configuration Model, \cite{Saracco2015}) represents the bipartite version of the Configuration Model, \cite{Park2004, Garlaschelli2008, Squartini2011}. The BiCM is defined by two degree sequences. Thus, our Hamiltonian is

\begin{equation}
H(\vec{\theta},\:\vec{C}(\mathbf{M}))=\sum_{r=1}^{N_R}\alpha_r k_r+\sum_{c=1}^{N_C}\beta_c h_c,
\end{equation}

\noindent where $k_r=\sum_{c=1}^{N_C}m_{rc}$ and $h_c=\sum_{r=1}^{N_R}m_{rc}$ are the degrees of nodes on the top and bottom layer, respectively; $\alpha_r$ and $\beta_c$, instead, are the Lagrangian multipliers associated with the constraints.

The probability of the generic matrix $\mathbf{M}$ thus reads

\beq\label{eq:pBiCMensemble}
P(\mathbf{M})=\dfrac{e^{-H(\vec{\theta},\:\vec{C}(\mathbf{M}))}}{Z(\vec{\theta})}
\eneq

\noindent where $Z(\vec{\theta})$ is the grand canonical partition function. It is possible to show that 

\begin{equation}\label{eq:pBiCMensemble2}
P(\mathbf{M})=\prod_{r=1}^{N_R}\prod_{c=1}^{N_C}p_{rc}^{m_{rc}}\left(1-p_{rc}\right)^{(1-m_{rc})},
\end{equation}

\noindent where 

\begin{equation}
p_{rc}=p_{rc}(\alpha_r,\beta_c)=\frac{e^{-(\alpha_r+\beta_c)}}{1+e^{-(\alpha_r+\beta_c)}}\equiv\frac{x_ry_c}{1+x_ry_c}
\label{bicmham}
\end{equation}

\noindent is the probability for a link between nodes $r$ and $c$ to exist.

In order to estimate the values for $x_r$ and $y_c$, let us maximize the probability of observing the given matrix $\mathbf{M}^*$, i.e. the likelihood function $\mathcal{L}=\text{ln}P(\mathbf{M}^*)$ \cite{Garlaschelli2008}. It is thus possible to derive the Lagrangian multipliers $\{x_r\}_{r=1}^{N_R}$ and $\{y_c\}_{c=1}^{N_C}$ by solving

\begin{equation}\label{eq:prcexplicit}
\begin{split}
k_r^*=\langle k_r\rangle=&\sum_{c=1}^{N_C}\dfrac{x_ry_c}{1+x_ry_c},\\
h_c^*=\langle h_c\rangle=&\sum_{r=1}^{N_R}\dfrac{x_ry_c}{1+x_ry_c}
\end{split}
\end{equation}
where $\{k_r^*\}_{r=1}^{N_R}$, $\{h_c^*\}_{c=1}^{N_C}$ are the observed degree sequences.

\subsection{Bipartite Partial Configuration Models}

Dealing with bipartite networks allows us to explore two ``partial'' versions of the BiCM (hereafter BiPCM), defined by constraining the degree sequences of, say, the top and bottom layer separately. Let us start with the null model $\text{BiPCM}_r$, defined by the following Hamiltonian:

\begin{equation}
H(\vec{\theta},\:\vec{C}(\mathbf{M}))=\sum_{r=1}^{N_R}\alpha_r k_r
\end{equation}

\noindent where $k_r=\sum_{c=1}^{N_C}m_{rc},\:\forall\:r$ are the degrees of nodes on the top layer. Although the probability of the generic matrix $\mathbf{M}$ still reads

\begin{equation}
P(\mathbf{M})=\prod_{r=1}^{N_R}\prod_{c=1}^{N_C}p_{rc}^{m_{rc}}\left(1-p_{rc}\right)^{(1-m_{rc})},
\end{equation}

\noindent upon ``switching off'' the multipliers $\{\beta_c\}_{c=1}^{N_C}$ the coefficient $p_{rc}$ now assumes the form

\begin{equation}
p_{rc}=p_{rc}(\alpha_r)=\frac{e^{-\alpha_r}}{1+e^{-\alpha_r}}\equiv\frac{x_r}{1+x_r},\:\forall\:c.
\label{bicmpart}
\end{equation}

Notice that the BiCM probability coefficients in (\ref{bicmham}) exactly reduce to the ones in (\ref{bicmpart}) whenever the degrees of all nodes belonging to the bottom layer coincide (i.e. $h_c\equiv h,\:\forall\:c$). However, $\text{BiPCM}_r$ provides an accurate approximation to the BiCM even when the values $\{h_c\}_{c=1}^{N_C}$ are characterized by a reduced degree of heterogeneity (e.g. as signalled by a coefficient of variation $c_v=s/m<0.5$, with $m$ and $s$ being, respectively, the mean and the standard deviation of the bottom layer degrees).

In order to estimate the values for $x_r$, let us maximize the likelihood function $\mathcal{L}=\text{ln}P(\mathbf{M}^*)$ again \cite{Garlaschelli2008}. It is thus possible to derive the Lagrangian multipliers $\{x_r\}_{r=1}^{N_R}$:

\begin{equation}
k_r^*=\langle k_r\rangle=\sum_{c=1}^{N_C}\dfrac{x_r}{1+x_r}\Longrightarrow p_{rc}=\frac{k_r^*}{N_C}.
\end{equation}

Notice that, in this case,

\begin{equation}
p(V^c_{rr'})=\frac{k_r^*k_{r'}^*}{N_C^2}
\end{equation}
i.e. each V-motif defined by $r$ and $r'$ has the same probability, independently from $c$. This, in turn, implies that the probability distribution of the number of V-motifs shared by nodes $r$ and $r'$ is again a Binomial distribution defined as

\begin{equation}
f_\text{Bin}(V_{rr'}=n)=\binom{N_C}{n}\left(\frac{k_r^*k_{r'}^*}{N_C^2}\right)^{n}\left(1-\frac{k_r^*k_{r'}^*}{N_C^2}\right)^{N_c-n}.
\end{equation}

Let us now move to considering the second partial null model, $\text{BiPCM}_c$, defined by the Hamiltonian

\begin{equation}
H(\vec{\theta},\:\vec{C}(\mathbf{M}))=\sum_{c=1}^{N_C}\beta_c h_c
\end{equation}

\noindent where $h_c=\sum_{r=1}^{N_R}m_{rc},\:\forall\:c$ are the degrees of nodes on the bottom layer. The probability of the generic matrix $\mathbf{M}$ still factorizes, with the coefficient $p_{rc}$ assuming the form

\begin{equation}
p_{rc}=p_{rc}(\beta_c)=\frac{e^{-\beta_c}}{1+e^{-\beta_c}}\equiv\frac{y_c}{1+y_c},\:\forall\:r;
\label{bicmpart2}
\end{equation}
as for the previously-considered $\text{BiPCM}$, the BiCM probability coefficients in (\ref{bicmham}) exactly reduce to the ones in (\ref{bicmpart2}) whenever the degrees of all nodes belonging to the top layer coincide (i.e. $k_r\equiv k,\:\forall\:r$). Again, when the values $\{k_r\}_{r=1}^{N_R}$ are characterized by a reduced degree of heterogeneity, $\text{BiPCM}_c$ provides an accurate approximation to the BiCM.

The Lagrangian multipliers $\{y_c\}_{c=1}^{N_C}$ are again straightforwardly estimated as

\begin{equation}
h_c^*=\langle h_c\rangle=\sum_{r=1}^{N_R}\dfrac{y_c}{1+y_c}\Longrightarrow p_{rc}=\frac{h_c^*}{N_R}.
\end{equation}

In this case, each V-motif defined by $r$, $r'$ and $c$ has a probability which depends exclusively on $c$. As a consequence, the probability distribution of the number of V-motifs shared by any two nodes $r$ and $r'$ is the same one, i.e. a Poisson-Binomial whose single Bernoulli trial is defined by a probability reading

\begin{equation}
p(V^c_{rr'})=\left(\frac{h_c^*}{N_R}\right)^2.
\end{equation}

\begin{figure}[t!]
\begin{center}
\includegraphics[width=0.49\textwidth]{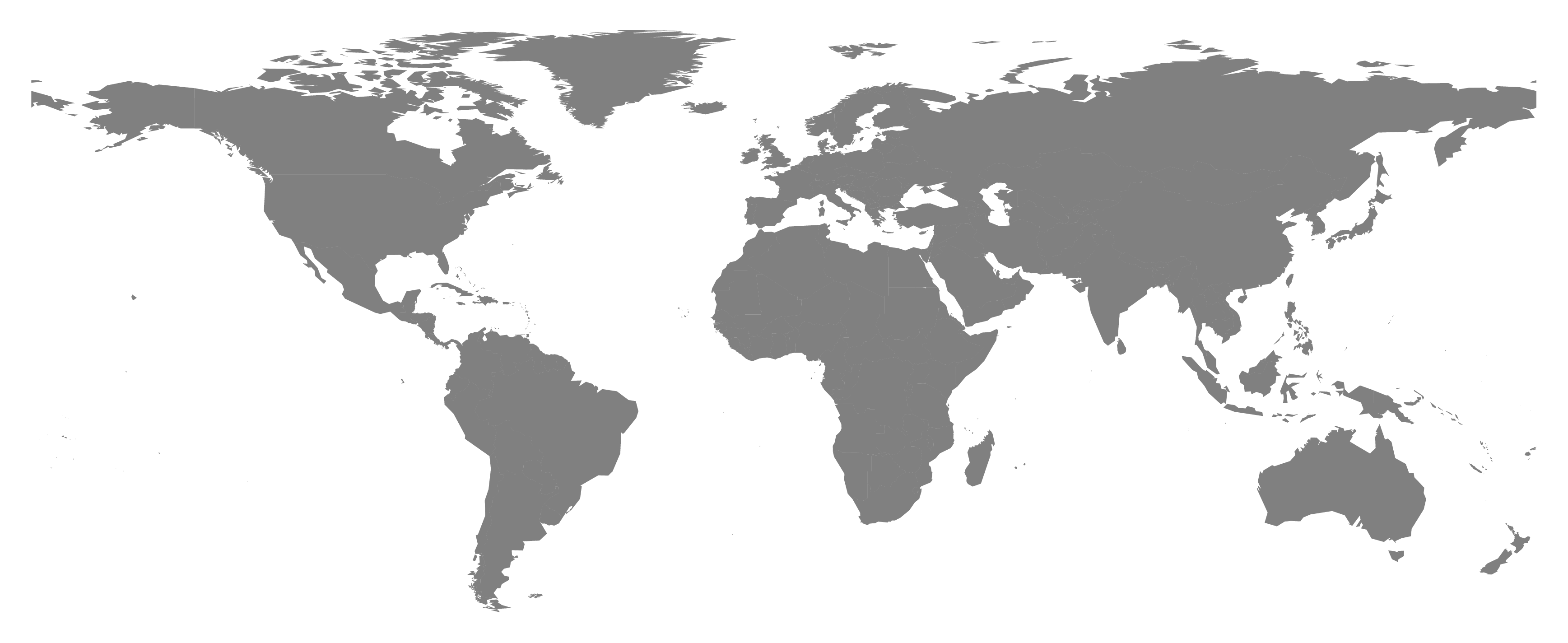}
\includegraphics[width=0.49\textwidth]{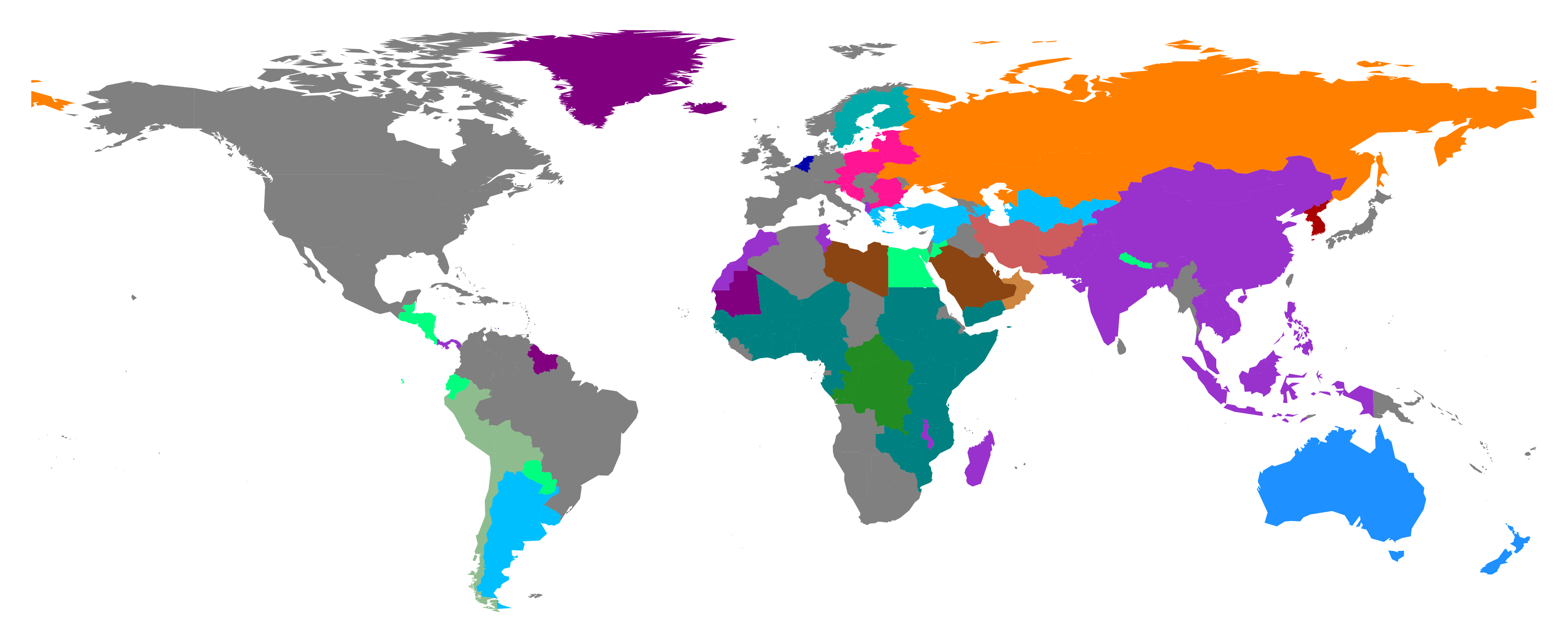}
\includegraphics[width=0.49\textwidth]{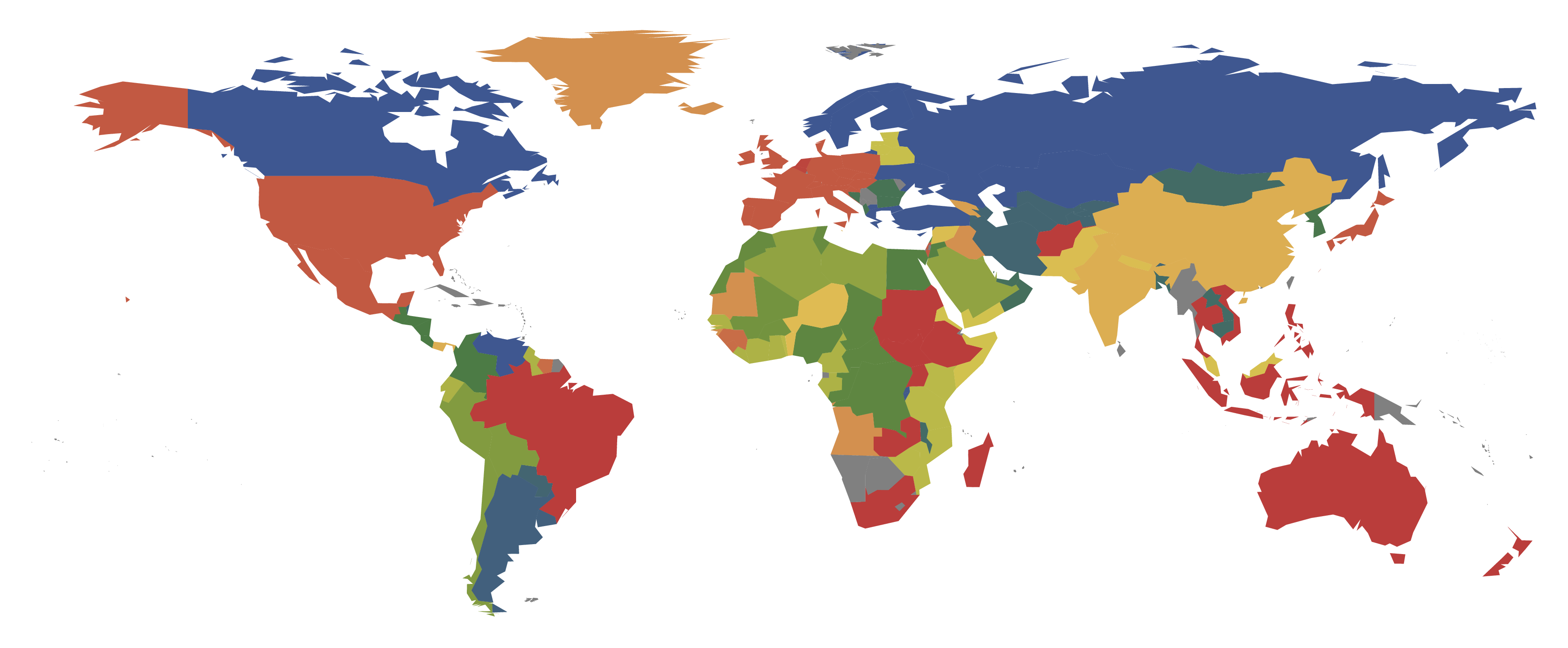}
\caption{Comparison between different projection methods, tested on the WTW in the year 2000. The method proposed in \cite{Tumminello2011} (top panel) outputs an empty projection: this may be due to the large number of hypotheses tested at a time, accounted for the Bonferroni correction. On the other hand, the links validated by the method proposed in \cite{Gualdi2016} (middle panel) constitute a subset of ours (as apparent by the partial overlap of the detected communities): in fact, applying the Bonferroni correction means selecting part of the links validated by FDR-controlling procedures. Last, links validated by the forest-inducing method proposed in \cite{Zaccaria2014} (bottom panel) are characterized by the largest overlap with the ones validated by our procedure ($\simeq 82\%$ - this large overlap may be due to the selection of those events having a high chance to be significant, even if an explicit control is missing).}
\label{fig:comparison}
\end{center}
\end{figure}

\section{comparing different projection algorithms}

Available procedures suffer from a number of limitations that our method aims at overcoming. In what follows we compare the performance of some of them in projecting the WTW on the countries layer, for the year 2000, in greater detail (see fig. \ref{fig:comparison} for the results of the comparison).

The method proposed in reference \cite{Tumminello2011} outputs an empty network for all years of our dataset: we suspect the reason to lie in the very large number of hypotheses tested at a time, leading to a too-severe correction. A similar result is obtained when applying the recipe proposed in \cite{Serrano2009}: only a tenth of links (among the group of advanced economies) are validated.

Although similar-in-spirit to ours, the method proposed in reference \cite{Gualdi2016} prescribes to implement the Bonferroni correction as well. All links validated by applying this kind of correction are always a subset of the links validated when controlling for the FDR: this is the reason underlying the less informative community structure obtained when this algorithm is run on the WTW.

The third comparison we have explicitly carried out is the one with the forest-inducing method proposed in reference \cite{Zaccaria2014}. Links validated by such a method are characterized by the largest overlap ($\simeq 82\%$) with the ones validated by our procedure. This may be due to the selection of those events which have the higher chance to be significant (i.e. the largest number of shared co-occurrences): anyway, no statistical control is explicitly provided (e.g. the forest-like topology is not {\it per se} guaranteed to encode the most significant events).
\newline
\newline
\indent As a final remark, we explicitly notice that the problem of spurious clustering does not affect our method, by definition. In fact, the presence of a node simultaneously connected to several nodes on the opposite layer does not imply the latter to be connected in the projection: this is the case if, and only if, the similarity between the involved nodes passes the test of statistical significance. An extreme example is provided by a network having a node $c$ (on one layer) which is connected to every other node (on the opposite layer), projected by employing the BiCM: since the fully-connected node is, actually, a ``deterministic'' node (its links are described by probability coefficients which are 1), any V-motif having it as a vertex (e.g. $V_{rr'}^c$) is deterministic as well. Thus, $P(V_{rr'}=0)=0$ (one V-motif is surely present) and the distribution describing the overlap between $r$ and $r'$ is shifted, as a whole, by one. In other words, the set of events which determine the presence of a link between $r$ and $r'$ does not include the deterministic V-motif (even more so, deterministic nodes can be discarded from the validation process carried out by the BiCM from the very beginning).

\section*{Acknowledgements}

This work was supported by the EU projects CoeGSS (grant num. 676547), Multiplex (grant num. 317532), Shakermaker (grant num. 687941), SoBigData (grant num. 654024) and the FET projects SIMPOL (grant num. 610704), DOLFINS (grant num. 640772). The authors acknowledge Alberto Cassese, Irene Grimaldi and all participants to NEDO Journal Club for useful discussions.

\bibliography{Bibliography}

\begin{thebibliography}{50}%
\makeatletter
\providecommand \@ifxundefined [1]{%
 \@ifx{#1\undefined}
}%
\providecommand \@ifnum [1]{%
 \ifnum #1\expandafter \@firstoftwo
 \else \expandafter \@secondoftwo
 \fi
}%
\providecommand \@ifx [1]{%
 \ifx #1\expandafter \@firstoftwo
 \else \expandafter \@secondoftwo
 \fi
}%
\providecommand \natexlab [1]{#1}%
\providecommand \enquote  [1]{``#1''}%
\providecommand \bibnamefont  [1]{#1}%
\providecommand \bibfnamefont [1]{#1}%
\providecommand \citenamefont [1]{#1}%
\providecommand \href@noop [0]{\@secondoftwo}%
\providecommand \href [0]{\begingroup \@sanitize@url \@href}%
\providecommand \@href[1]{\@@startlink{#1}\@@href}%
\providecommand \@@href[1]{\endgroup#1\@@endlink}%
\providecommand \@sanitize@url [0]{\catcode `\\12\catcode `\$12\catcode
  `\&12\catcode `\#12\catcode `\^12\catcode `\_12\catcode `\%12\relax}%
\providecommand \@@startlink[1]{}%
\providecommand \@@endlink[0]{}%
\providecommand \url  [0]{\begingroup\@sanitize@url \@url }%
\providecommand \@url [1]{\endgroup\@href {#1}{\urlprefix }}%
\providecommand \urlprefix  [0]{URL }%
\providecommand \Eprint [0]{\href }%
\providecommand \doibase [0]{http://dx.doi.org/}%
\providecommand \selectlanguage [0]{\@gobble}%
\providecommand \bibinfo  [0]{\@secondoftwo}%
\providecommand \bibfield  [0]{\@secondoftwo}%
\providecommand \translation [1]{[#1]}%
\providecommand \BibitemOpen [0]{}%
\providecommand \bibitemStop [0]{}%
\providecommand \bibitemNoStop [0]{.\EOS\space}%
\providecommand \EOS [0]{\spacefactor3000\relax}%
\providecommand \BibitemShut  [1]{\csname bibitem#1\endcsname}%
\let\auto@bib@innerbib\@empty
\bibitem [{\citenamefont {Caldarelli}(2007)}]{Caldarelli2007}%
  \BibitemOpen
  \bibfield  {author} {\bibinfo {author} {\bibfnamefont {G.}~\bibnamefont
  {Caldarelli}},\ }\href@noop {} {\emph {\bibinfo {title} {Scale-free Networks:
  Complex Webs in Nature and Technology}}}\ (\bibinfo  {publisher} {Oxford
  University Press},\ \bibinfo {year} {2007})\BibitemShut {NoStop}%
\bibitem [{\citenamefont {Newman}(2010)}]{Newman2010}%
  \BibitemOpen
  \bibfield  {author} {\bibinfo {author} {\bibfnamefont {M.}~\bibnamefont
  {Newman}},\ }\href@noop {} {\emph {\bibinfo {title} {Networks: An
  Introduction}}}\ (\bibinfo  {publisher} {Oxford University Press},\ \bibinfo
  {year} {2010})\BibitemShut {NoStop}%
\bibitem [{\citenamefont {Neal}(2014)}]{Review2014}%
  \BibitemOpen
  \bibfield  {author} {\bibinfo {author} {\bibfnamefont {Z.}~\bibnamefont
  {Neal}},\ }\href {\doibase 10.1016/j.socnet.2014.06.001} {\bibfield
  {journal} {\bibinfo  {journal} {Social Networks}\ }\textbf {\bibinfo {volume}
  {39}},\ \bibinfo {pages} {84} (\bibinfo {year} {2014})}\BibitemShut {NoStop}%
\bibitem [{\citenamefont {Latapy}\ \emph {et~al.}(2008)\citenamefont {Latapy},
  \citenamefont {Magnien},\ and\ \citenamefont {del Vecchio}}]{Latapy2008}%
  \BibitemOpen
  \bibfield  {author} {\bibinfo {author} {\bibfnamefont {M.}~\bibnamefont
  {Latapy}}, \bibinfo {author} {\bibfnamefont {C.}~\bibnamefont {Magnien}}, \
  and\ \bibinfo {author} {\bibfnamefont {N.}~\bibnamefont {del Vecchio}},\
  }\href {\doibase 10.1016/j.socnet.2007.04.006} {\bibfield  {journal}
  {\bibinfo  {journal} {Social Networks}\ }\textbf {\bibinfo {volume} {30}},\
  \bibinfo {pages} {31} (\bibinfo {year} {2008})}\BibitemShut {NoStop}%
\bibitem [{\citenamefont {Watts}\ and\ \citenamefont
  {Strogatz}(1998)}]{Watts1998}%
  \BibitemOpen
  \bibfield  {author} {\bibinfo {author} {\bibfnamefont {D.~J.}\ \bibnamefont
  {Watts}}\ and\ \bibinfo {author} {\bibfnamefont {S.~H.}\ \bibnamefont
  {Strogatz}},\ }\href {\doibase 10.1038/30918} {\bibfield  {journal} {\bibinfo
   {journal} {Nature}\ }\textbf {\bibinfo {volume} {393}},\ \bibinfo {pages}
  {440} (\bibinfo {year} {1998})}\BibitemShut {NoStop}%
\bibitem [{\citenamefont {Derudder}\ and\ \citenamefont
  {Taylor}(2005)}]{Derudder2005}%
  \BibitemOpen
  \bibfield  {author} {\bibinfo {author} {\bibfnamefont {B.}~\bibnamefont
  {Derudder}}\ and\ \bibinfo {author} {\bibfnamefont {P.}~\bibnamefont
  {Taylor}},\ }\href {\doibase 10.1111/j.1471-0374.2005.00108.x} {\bibfield
  {journal} {\bibinfo  {journal} {Global Networks}\ }\textbf {\bibinfo {volume}
  {5}},\ \bibinfo {pages} {71} (\bibinfo {year} {2005})}\BibitemShut {NoStop}%
\bibitem [{\citenamefont {Serrano}\ \emph {et~al.}(2009)\citenamefont
  {Serrano}, \citenamefont {Bogun{\'a}},\ and\ \citenamefont
  {Vespignani}}]{Serrano2009}%
  \BibitemOpen
  \bibfield  {author} {\bibinfo {author} {\bibfnamefont {M.~A.}\ \bibnamefont
  {Serrano}}, \bibinfo {author} {\bibfnamefont {M.}~\bibnamefont {Bogun{\'a}}},
  \ and\ \bibinfo {author} {\bibfnamefont {A.}~\bibnamefont {Vespignani}},\
  }\href {\doibase 10.1073/pnas.0808904106} {\bibfield  {journal} {\bibinfo
  {journal} {Proceedings of the National Academy of Sciences}\ }\textbf
  {\bibinfo {volume} {106}},\ \bibinfo {pages} {6483} (\bibinfo {year}
  {2009})}\BibitemShut {NoStop}%
\bibitem [{\citenamefont {Caldarelli}\ \emph {et~al.}(2012)\citenamefont
  {Caldarelli}, \citenamefont {Cristelli}, \citenamefont {Gabrielli},
  \citenamefont {Pietronero}, \citenamefont {Scala},\ and\ \citenamefont
  {Tacchella}}]{Caldarelli2012}%
  \BibitemOpen
  \bibfield  {author} {\bibinfo {author} {\bibfnamefont {G.}~\bibnamefont
  {Caldarelli}}, \bibinfo {author} {\bibfnamefont {M.}~\bibnamefont
  {Cristelli}}, \bibinfo {author} {\bibfnamefont {A.}~\bibnamefont
  {Gabrielli}}, \bibinfo {author} {\bibfnamefont {L.}~\bibnamefont
  {Pietronero}}, \bibinfo {author} {\bibfnamefont {A.}~\bibnamefont {Scala}}, \
  and\ \bibinfo {author} {\bibfnamefont {A.}~\bibnamefont {Tacchella}},\ }\href
  {\doibase 10.1371/journal.pone.0047278} {\bibfield  {journal} {\bibinfo
  {journal} {PLoS ONE}\ }\textbf {\bibinfo {volume} {7}},\ \bibinfo {pages}
  {e47278} (\bibinfo {year} {2012})}\BibitemShut {NoStop}%
\bibitem [{\citenamefont {Zaccaria}\ \emph {et~al.}(2014)\citenamefont
  {Zaccaria}, \citenamefont {Cristelli}, \citenamefont {Tacchella},\ and\
  \citenamefont {Pietronero}}]{Zaccaria2014}%
  \BibitemOpen
  \bibfield  {author} {\bibinfo {author} {\bibfnamefont {A.}~\bibnamefont
  {Zaccaria}}, \bibinfo {author} {\bibfnamefont {M.}~\bibnamefont {Cristelli}},
  \bibinfo {author} {\bibfnamefont {A.}~\bibnamefont {Tacchella}}, \ and\
  \bibinfo {author} {\bibfnamefont {L.}~\bibnamefont {Pietronero}},\ }\href
  {\doibase 10.1371/journal.pone.0113770} {\bibfield  {journal} {\bibinfo
  {journal} {PLoS ONE}\ }\textbf {\bibinfo {volume} {9}},\ \bibinfo {pages}
  {e113770} (\bibinfo {year} {2014})}\BibitemShut {NoStop}%
\bibitem [{\citenamefont {Linden}\ \emph {et~al.}(2003)\citenamefont {Linden},
  \citenamefont {Smith},\ and\ \citenamefont {York}}]{Amazon.com}%
  \BibitemOpen
  \bibfield  {author} {\bibinfo {author} {\bibfnamefont {G.}~\bibnamefont
  {Linden}}, \bibinfo {author} {\bibfnamefont {B.}~\bibnamefont {Smith}}, \
  and\ \bibinfo {author} {\bibfnamefont {J.}~\bibnamefont {York}},\ }\href
  {\doibase 10.1109/MIC.2003.1167344} {\bibfield  {journal} {\bibinfo
  {journal} {IEEE Internet Computing}\ }\textbf {\bibinfo {volume} {7}},\
  \bibinfo {pages} {76} (\bibinfo {year} {2003})}\BibitemShut {NoStop}%
\bibitem [{\citenamefont {Bonacich}(1972)}]{Bonacich1972}%
  \BibitemOpen
  \bibfield  {author} {\bibinfo {author} {\bibfnamefont {P.}~\bibnamefont
  {Bonacich}},\ }\href@noop {} {\bibfield  {journal} {\bibinfo  {journal}
  {Social Methodologies}\ }\textbf {\bibinfo {volume} {4}},\ \bibinfo {pages}
  {176} (\bibinfo {year} {1972})}\BibitemShut {NoStop}%
\bibitem [{\citenamefont {{Tumminello}}\ \emph {et~al.}(2011)\citenamefont
  {{Tumminello}}, \citenamefont {{Miccich{\'e}}}, \citenamefont {{Lillo}},
  \citenamefont {{Piilo}},\ and\ \citenamefont {{Mantegna}}}]{Tumminello2011}%
  \BibitemOpen
  \bibfield  {author} {\bibinfo {author} {\bibfnamefont {M.}~\bibnamefont
  {{Tumminello}}}, \bibinfo {author} {\bibfnamefont {S.}~\bibnamefont
  {{Miccich{\'e}}}}, \bibinfo {author} {\bibfnamefont {F.}~\bibnamefont
  {{Lillo}}}, \bibinfo {author} {\bibfnamefont {J.}~\bibnamefont {{Piilo}}}, \
  and\ \bibinfo {author} {\bibfnamefont {R.~N.}\ \bibnamefont {{Mantegna}}},\
  }\href {\doibase 10.1371/journal.pone.0017994} {\bibfield  {journal}
  {\bibinfo  {journal} {PLoS ONE}\ }\textbf {\bibinfo {volume} {6}},\ \bibinfo
  {pages} {e17994} (\bibinfo {year} {2011})}\BibitemShut {NoStop}%
\bibitem [{\citenamefont {Gualdi}\ \emph {et~al.}(2016)\citenamefont {Gualdi},
  \citenamefont {Cimini}, \citenamefont {Primicerio}, \citenamefont
  {Di~Clemente},\ and\ \citenamefont {Challet}}]{Gualdi2016}%
  \BibitemOpen
  \bibfield  {author} {\bibinfo {author} {\bibfnamefont {S.}~\bibnamefont
  {Gualdi}}, \bibinfo {author} {\bibfnamefont {G.}~\bibnamefont {Cimini}},
  \bibinfo {author} {\bibfnamefont {K.}~\bibnamefont {Primicerio}}, \bibinfo
  {author} {\bibfnamefont {R.}~\bibnamefont {Di~Clemente}}, \ and\ \bibinfo
  {author} {\bibfnamefont {D.}~\bibnamefont {Challet}},\ }\href
  {https://arxiv.org/abs/1603.05914} {\bibfield  {journal} {\bibinfo  {journal}
  {arXiv preprint arXiv:1603.05914}\ } (\bibinfo {year} {2016})}\BibitemShut
  {NoStop}%
\bibitem [{\citenamefont {Dianati}(2016)}]{Dianati2016}%
  \BibitemOpen
  \bibfield  {author} {\bibinfo {author} {\bibfnamefont {N.}~\bibnamefont
  {Dianati}},\ }\href {https://arxiv.org/abs/1607.01735} {\bibfield  {journal}
  {\bibinfo  {journal} {arXiv preprint arXiv:1607.01735}\ } (\bibinfo {year}
  {2016})}\BibitemShut {NoStop}%
\bibitem [{\citenamefont {Guillaume}\ and\ \citenamefont
  {Latapy}(2003)}]{Guillaume2003}%
  \BibitemOpen
  \bibfield  {author} {\bibinfo {author} {\bibfnamefont {J.-L.}\ \bibnamefont
  {Guillaume}}\ and\ \bibinfo {author} {\bibfnamefont {M.}~\bibnamefont
  {Latapy}},\ }\href {https://arxiv.org/abs/cond-mat/0307095} {\bibfield
  {journal} {\bibinfo  {journal} {arXiv preprint arXiv:cond-mat/0307095}\ }
  (\bibinfo {year} {2003})}\BibitemShut {NoStop}%
\bibitem [{\citenamefont {Saracco}\ \emph {et~al.}(2015)\citenamefont
  {Saracco}, \citenamefont {Di~Clemente}, \citenamefont {Gabrielli},\ and\
  \citenamefont {Squartini}}]{Saracco2015}%
  \BibitemOpen
  \bibfield  {author} {\bibinfo {author} {\bibfnamefont {F.}~\bibnamefont
  {Saracco}}, \bibinfo {author} {\bibfnamefont {R.}~\bibnamefont
  {Di~Clemente}}, \bibinfo {author} {\bibfnamefont {A.}~\bibnamefont
  {Gabrielli}}, \ and\ \bibinfo {author} {\bibfnamefont {T.}~\bibnamefont
  {Squartini}},\ }\href {\doibase 10.1038/srep10595} {\bibfield  {journal}
  {\bibinfo  {journal} {Scientific Reports}\ }\textbf {\bibinfo {volume} {5}},\
  \bibinfo {pages} {10595} (\bibinfo {year} {2015})}\BibitemShut {NoStop}%
\bibitem [{\citenamefont {Diestel}(2006)}]{Diestel2006}%
  \BibitemOpen
  \bibfield  {author} {\bibinfo {author} {\bibfnamefont {R.}~\bibnamefont
  {Diestel}},\ }\href@noop {} {\emph {\bibinfo {title} {Graph Theory}}}\
  (\bibinfo  {publisher} {Springer-Verlag Heidelberg, New York (USA)},\
  \bibinfo {year} {2006})\BibitemShut {NoStop}%
\bibitem [{\citenamefont {Park}\ and\ \citenamefont {Newman}(2004)}]{Park2004}%
  \BibitemOpen
  \bibfield  {author} {\bibinfo {author} {\bibfnamefont {J.}~\bibnamefont
  {Park}}\ and\ \bibinfo {author} {\bibfnamefont {M.~E.~J.}\ \bibnamefont
  {Newman}},\ }\href {\doibase 10.1103/PhysRevE.70.066117} {\bibfield
  {journal} {\bibinfo  {journal} {Physical Review E}\ }\textbf {\bibinfo
  {volume} {70}},\ \bibinfo {pages} {066117} (\bibinfo {year}
  {2004})}\BibitemShut {NoStop}%
\bibitem [{\citenamefont {Garlaschelli}\ and\ \citenamefont
  {Loffredo}(2008)}]{Garlaschelli2008}%
  \BibitemOpen
  \bibfield  {author} {\bibinfo {author} {\bibfnamefont {D.}~\bibnamefont
  {Garlaschelli}}\ and\ \bibinfo {author} {\bibfnamefont {M.~I.}\ \bibnamefont
  {Loffredo}},\ }\href {\doibase 10.1103/PhysRevE.78.015101} {\bibfield
  {journal} {\bibinfo  {journal} {Physical Review E}\ }\textbf {\bibinfo
  {volume} {78}},\ \bibinfo {pages} {015101} (\bibinfo {year}
  {2008})}\BibitemShut {NoStop}%
\bibitem [{\citenamefont {{Squartini}}\ and\ \citenamefont
  {{Garlaschelli}}(2011)}]{Squartini2011}%
  \BibitemOpen
  \bibfield  {author} {\bibinfo {author} {\bibfnamefont {T.}~\bibnamefont
  {{Squartini}}}\ and\ \bibinfo {author} {\bibfnamefont {D.}~\bibnamefont
  {{Garlaschelli}}},\ }\href {\doibase 10.1088/1367-2630/13/8/083001}
  {\bibfield  {journal} {\bibinfo  {journal} {New Journal of Physics}\ }\textbf
  {\bibinfo {volume} {13}},\ \bibinfo {pages} {083001} (\bibinfo {year}
  {2011})}\BibitemShut {NoStop}%
\bibitem [{\citenamefont {Fronczak}(2014)}]{Fronczak2012}%
  \BibitemOpen
  \bibfield  {author} {\bibinfo {author} {\bibfnamefont {A.}~\bibnamefont
  {Fronczak}},\ }\href {\doibase 10.1007/978-1-4614-6170-8} {\bibfield
  {journal} {\bibinfo  {journal} {Encyclopedia of Social Network Analysis and
  Mining}\ ,\ \bibinfo {pages} {500}} (\bibinfo {year} {2014})}\BibitemShut
  {NoStop}%
\bibitem [{\citenamefont {{Mastrandrea}}\ \emph {et~al.}(2014)\citenamefont
  {{Mastrandrea}}, \citenamefont {{Squartini}}, \citenamefont {{Fagiolo}},\
  and\ \citenamefont {{Garlaschelli}}}]{Mastrandrea2014}%
  \BibitemOpen
  \bibfield  {author} {\bibinfo {author} {\bibfnamefont {R.}~\bibnamefont
  {{Mastrandrea}}}, \bibinfo {author} {\bibfnamefont {T.}~\bibnamefont
  {{Squartini}}}, \bibinfo {author} {\bibfnamefont {G.}~\bibnamefont
  {{Fagiolo}}}, \ and\ \bibinfo {author} {\bibfnamefont {D.}~\bibnamefont
  {{Garlaschelli}}},\ }\href {\doibase 10.1088/1367-2630/16/4/043022}
  {\bibfield  {journal} {\bibinfo  {journal} {New Journal of Physics}\ }\textbf
  {\bibinfo {volume} {16}},\ \bibinfo {pages} {043022} (\bibinfo {year}
  {2014})}\BibitemShut {NoStop}%
\bibitem [{\citenamefont {Hoeffding}(1956)}]{hoeffding1956}%
  \BibitemOpen
  \bibfield  {author} {\bibinfo {author} {\bibfnamefont {W.}~\bibnamefont
  {Hoeffding}},\ }\href {\doibase 10.1214/aoms/1177728178} {\bibfield
  {journal} {\bibinfo  {journal} {The Annals of Mathematical Statistics}\
  }\textbf {\bibinfo {volume} {27}},\ \bibinfo {pages} {713} (\bibinfo {year}
  {1956})}\BibitemShut {NoStop}%
\bibitem [{\citenamefont {Wang}(1993)}]{wang1993}%
  \BibitemOpen
  \bibfield  {author} {\bibinfo {author} {\bibfnamefont {Y.~H.}\ \bibnamefont
  {Wang}},\ }\href {http://www.jstor.org/stable/24304959} {\bibfield  {journal}
  {\bibinfo  {journal} {Statistica Sinica}\ }\textbf {\bibinfo {volume} {3}},\
  \bibinfo {pages} {295} (\bibinfo {year} {1993})}\BibitemShut {NoStop}%
\bibitem [{Note1()}]{Note1}%
  \BibitemOpen
  \bibinfo {note} {Python code for computing p-values under the null models
  discussed in the paper: \protect \url
  {https://github.com/tsakim/bicm}}\BibitemShut {NoStop}%
\bibitem [{\citenamefont {Benjamini}\ and\ \citenamefont
  {Hochberg}(1995)}]{Benjamini1995}%
  \BibitemOpen
  \bibfield  {author} {\bibinfo {author} {\bibfnamefont {Y.}~\bibnamefont
  {Benjamini}}\ and\ \bibinfo {author} {\bibfnamefont {Y.}~\bibnamefont
  {Hochberg}},\ }\href@noop {} {\bibfield  {journal} {\bibinfo  {journal}
  {Journal of the Royal Statistical Society B}\ }\textbf {\bibinfo {volume}
  {57}},\ \bibinfo {pages} {289} (\bibinfo {year} {1995})}\BibitemShut
  {NoStop}%
\bibitem [{\citenamefont {Blondel}\ \emph {et~al.}(2008)\citenamefont
  {Blondel}, \citenamefont {Guillaume}, \citenamefont {Lambiotte},\ and\
  \citenamefont {Lefebvre}}]{Blondel2008}%
  \BibitemOpen
  \bibfield  {author} {\bibinfo {author} {\bibfnamefont {V.~D.}\ \bibnamefont
  {Blondel}}, \bibinfo {author} {\bibfnamefont {J.-L.}\ \bibnamefont
  {Guillaume}}, \bibinfo {author} {\bibfnamefont {R.}~\bibnamefont
  {Lambiotte}}, \ and\ \bibinfo {author} {\bibfnamefont {E.}~\bibnamefont
  {Lefebvre}},\ }\href {\doibase 10.1088/1742-5468/2008/10/P10008} {\bibfield
  {journal} {\bibinfo  {journal} {Journal of Statistical Mechanics: Theory and
  Experiment}\ }\textbf {\bibinfo {volume} {2008}},\ \bibinfo {pages} {P10008}
  (\bibinfo {year} {2008})}\BibitemShut {NoStop}%
\bibitem [{\citenamefont {{Fortunato}}(2010)}]{Fortunato2010}%
  \BibitemOpen
  \bibfield  {author} {\bibinfo {author} {\bibfnamefont {S.}~\bibnamefont
  {{Fortunato}}},\ }\href {\doibase 10.1016/j.physrep.2009.11.002} {\bibfield
  {journal} {\bibinfo  {journal} {Physics Reports}\ }\textbf {\bibinfo {volume}
  {486}},\ \bibinfo {pages} {75} (\bibinfo {year} {2010})}\BibitemShut
  {NoStop}%
\bibitem [{\citenamefont {Staudt}\ and\ \citenamefont
  {Meyerhenke}(2016)}]{staudt2016engineering}%
  \BibitemOpen
  \bibfield  {author} {\bibinfo {author} {\bibfnamefont {C.~L.}\ \bibnamefont
  {Staudt}}\ and\ \bibinfo {author} {\bibfnamefont {H.}~\bibnamefont
  {Meyerhenke}},\ }\href@noop {} {\bibfield  {journal} {\bibinfo  {journal}
  {IEEE Transactions on Parallel and Distributed Systems}\ }\textbf {\bibinfo
  {volume} {27}},\ \bibinfo {pages} {171} (\bibinfo {year} {2016})}\BibitemShut
  {NoStop}%
\bibitem [{Note2()}]{Note2}%
  \BibitemOpen
  \bibinfo {note} {\protect \url
  {https://github.com/tsakim/Shuffled_Louvain}.}\BibitemShut {Stop}%
\bibitem [{Note3()}]{Note3}%
  \BibitemOpen
  \bibinfo {note} {\protect \url {http://comtrade.un.org/}}\BibitemShut
  {NoStop}%
\bibitem [{\citenamefont {G.~Gaulier}(2013)}]{BACI2013}%
  \BibitemOpen
  \bibfield  {author} {\bibinfo {author} {\bibfnamefont {S.}~\bibnamefont
  {G.~Gaulier}},\ }\href {http://www.cepii.fr/anglaisgraph/workpap/pdf/2010/
  wp2010-23.pdf} {\bibfield  {journal} {\bibinfo  {journal} {BACI:
  International Trade Database at the Product Level}\ } (\bibinfo {year} {last
  access: 5 July 2013})}\BibitemShut {NoStop}%
\bibitem [{\citenamefont {Tacchella}\ \emph {et~al.}(2012)\citenamefont
  {Tacchella}, \citenamefont {Cristelli}, \citenamefont {Caldarelli},
  \citenamefont {Gabrielli},\ and\ \citenamefont {Pietronero}}]{Tacchella2012}%
  \BibitemOpen
  \bibfield  {author} {\bibinfo {author} {\bibfnamefont {A.}~\bibnamefont
  {Tacchella}}, \bibinfo {author} {\bibfnamefont {M.}~\bibnamefont
  {Cristelli}}, \bibinfo {author} {\bibfnamefont {G.}~\bibnamefont
  {Caldarelli}}, \bibinfo {author} {\bibfnamefont {A.}~\bibnamefont
  {Gabrielli}}, \ and\ \bibinfo {author} {\bibfnamefont {L.}~\bibnamefont
  {Pietronero}},\ }\href {http://dx.doi.org/10.1038/srep00723} {\bibfield
  {journal} {\bibinfo  {journal} {Scientific Reports}\ }\textbf {\bibinfo
  {volume} {2}},\ \bibinfo {pages} {723} (\bibinfo {year} {2012})}\BibitemShut
  {NoStop}%
\bibitem [{\citenamefont {Di~Clemente}\ \emph {et~al.}(2014)\citenamefont
  {Di~Clemente}, \citenamefont {Chiarotti}, \citenamefont {Cristelli},
  \citenamefont {Tacchella},\ and\ \citenamefont
  {Pietronero}}]{DiClemente2014}%
  \BibitemOpen
  \bibfield  {author} {\bibinfo {author} {\bibfnamefont {R.}~\bibnamefont
  {Di~Clemente}}, \bibinfo {author} {\bibfnamefont {G.~L.}\ \bibnamefont
  {Chiarotti}}, \bibinfo {author} {\bibfnamefont {M.}~\bibnamefont
  {Cristelli}}, \bibinfo {author} {\bibfnamefont {A.}~\bibnamefont
  {Tacchella}}, \ and\ \bibinfo {author} {\bibfnamefont {L.}~\bibnamefont
  {Pietronero}},\ }\href {\doibase 10.1371/journal.pone.0112525} {\bibfield
  {journal} {\bibinfo  {journal} {PLoS ONE}\ }\textbf {\bibinfo {volume} {9}},\
  \bibinfo {pages} {e112525} (\bibinfo {year} {2014})}\BibitemShut {NoStop}%
\bibitem [{\citenamefont {Hidalgo}\ \emph {et~al.}(2007)\citenamefont
  {Hidalgo}, \citenamefont {Klinger}, \citenamefont {Barab{\'a}si},\ and\
  \citenamefont {Hausmann}}]{Hidalgo2007}%
  \BibitemOpen
  \bibfield  {author} {\bibinfo {author} {\bibfnamefont {C.~A.}\ \bibnamefont
  {Hidalgo}}, \bibinfo {author} {\bibfnamefont {B.}~\bibnamefont {Klinger}},
  \bibinfo {author} {\bibfnamefont {A.-L.}\ \bibnamefont {Barab{\'a}si}}, \
  and\ \bibinfo {author} {\bibfnamefont {R.}~\bibnamefont {Hausmann}},\ }\href
  {\doibase 10.1126/science.1144581} {\bibfield  {journal} {\bibinfo  {journal}
  {Science}\ }\textbf {\bibinfo {volume} {317}},\ \bibinfo {pages} {482}
  (\bibinfo {year} {2007})}\BibitemShut {NoStop}%
\bibitem [{\citenamefont {Hidalgo}\ and\ \citenamefont
  {Hausmann}(2009)}]{Hidalgo2009}%
  \BibitemOpen
  \bibfield  {author} {\bibinfo {author} {\bibfnamefont {C.~A.}\ \bibnamefont
  {Hidalgo}}\ and\ \bibinfo {author} {\bibfnamefont {R.}~\bibnamefont
  {Hausmann}},\ }\href {\doibase 10.1073/pnas.0900943106} {\bibfield  {journal}
  {\bibinfo  {journal} {Proceedings of the National Academy of Sciences}\
  }\textbf {\bibinfo {volume} {106}},\ \bibinfo {pages} {10570} (\bibinfo
  {year} {2009})}\BibitemShut {NoStop}%
\bibitem [{\citenamefont {Hausmann}\ and\ \citenamefont
  {Hidalgo}(2010)}]{Hausmann2010}%
  \BibitemOpen
  \bibfield  {author} {\bibinfo {author} {\bibfnamefont {R.}~\bibnamefont
  {Hausmann}}\ and\ \bibinfo {author} {\bibfnamefont {C.}~\bibnamefont
  {Hidalgo}},\ }\href@noop {} {\bibfield  {journal} {\bibinfo  {journal} {HKS
  Working Paper No. RWP 10-045}\ } (\bibinfo {year} {2010})}\BibitemShut
  {NoStop}%
\bibitem [{\citenamefont {Cristelli}\ \emph {et~al.}(2013)\citenamefont
  {Cristelli}, \citenamefont {Gabrielli}, \citenamefont {Tacchella},
  \citenamefont {Caldarelli},\ and\ \citenamefont
  {Pietronero}}]{Cristelli2013}%
  \BibitemOpen
  \bibfield  {author} {\bibinfo {author} {\bibfnamefont {M.}~\bibnamefont
  {Cristelli}}, \bibinfo {author} {\bibfnamefont {A.}~\bibnamefont
  {Gabrielli}}, \bibinfo {author} {\bibfnamefont {A.}~\bibnamefont
  {Tacchella}}, \bibinfo {author} {\bibfnamefont {G.}~\bibnamefont
  {Caldarelli}}, \ and\ \bibinfo {author} {\bibfnamefont {L.}~\bibnamefont
  {Pietronero}},\ }\href {\doibase 10.1371/journal.pone.0070726} {\bibfield
  {journal} {\bibinfo  {journal} {PLoS ONE}\ }\textbf {\bibinfo {volume} {8}},\
  \bibinfo {pages} {e70726} (\bibinfo {year} {2013})}\BibitemShut {NoStop}%
\bibitem [{\citenamefont {Tacchella}\ \emph {et~al.}(2013)\citenamefont
  {Tacchella}, \citenamefont {Cristelli}, \citenamefont {Caldarelli},
  \citenamefont {Gabrielli},\ and\ \citenamefont {Pietronero}}]{Tacchella2013}%
  \BibitemOpen
  \bibfield  {author} {\bibinfo {author} {\bibfnamefont {A.}~\bibnamefont
  {Tacchella}}, \bibinfo {author} {\bibfnamefont {M.}~\bibnamefont
  {Cristelli}}, \bibinfo {author} {\bibfnamefont {G.}~\bibnamefont
  {Caldarelli}}, \bibinfo {author} {\bibfnamefont {A.}~\bibnamefont
  {Gabrielli}}, \ and\ \bibinfo {author} {\bibfnamefont {L.}~\bibnamefont
  {Pietronero}},\ }\href@noop {} {\bibfield  {journal} {\bibinfo  {journal}
  {Journal of Economic Dynamics and Control}\ }\textbf {\bibinfo {volume}
  {37}},\ \bibinfo {pages} {1683} (\bibinfo {year} {2013})}\BibitemShut
  {NoStop}%
\bibitem [{\citenamefont {Straka}\ \emph {et~al.}(2016)\citenamefont {Straka},
  \citenamefont {Saracco},\ and\ \citenamefont {Caldarelli}}]{procmik}%
  \BibitemOpen
  \bibfield  {author} {\bibinfo {author} {\bibfnamefont {M.~J.}\ \bibnamefont
  {Straka}}, \bibinfo {author} {\bibfnamefont {F.}~\bibnamefont {Saracco}}, \
  and\ \bibinfo {author} {\bibfnamefont {G.}~\bibnamefont {Caldarelli}},\ }in\
  \href {http://www.complexnetworks.org/BookOfAbstractsCNA16.pdf} {\emph
  {\bibinfo {booktitle} {Product Similarities in International Trade from
  Entropy-based Null Models}}}\ (\bibinfo {year} {2016})\ pp.\ \bibinfo {pages}
  {130--132}\BibitemShut {NoStop}%
\bibitem [{Note4()}]{Note4}%
  \BibitemOpen
  \bibinfo {note} {``Cow'' by Nook Fulloption; ``Fish'' by Iconic;
  ``Excavator'' by Kokota; ``Light bulb'' by Hopkins; ``Milk'' by Artem
  Kovyazin; ``Curved Pipe'' by Oliviu Stoian; ``Tractor'' by Iconic;
  ``Recycle'' by Agus Purwanto; ``Experiment'' by Made by Made; ``Accumulator''
  by Aleksandr Vector; ``Washing Machine'' by Tomas Knopp; ``Metal'' by Leif
  Michelsen; ``Screw'' by Creaticca Creative Agency; ``Tram'' by Gleb
  Khorunzhiy; ``Turbine'' by Luigi Di Capua; ``Tire'' by Rediffusion; ``Ball Of
  Yarn'' by Denis Sazhin; ``Fabric'' by Oliviu Stoian; ``Shoe'' by Giuditta
  Valentina Gentile; ``Clothing'' by Marvdrock; ``Candies'' by Creative Mania;
  ``Wood Plank'' by Cono Studio Milano; ``Wood Logs'' by Alice Noir from the
  Noun Project. All icons are under the CC licence.}\BibitemShut {Stop}%
\bibitem [{\citenamefont {Resnick}\ \emph {et~al.}(1994)\citenamefont
  {Resnick}, \citenamefont {Iacovou}, \citenamefont {Suchak}, \citenamefont
  {Bergstrom},\ and\ \citenamefont {Riedl}}]{GroupLens}%
  \BibitemOpen
  \bibfield  {author} {\bibinfo {author} {\bibfnamefont {P.}~\bibnamefont
  {Resnick}}, \bibinfo {author} {\bibfnamefont {N.}~\bibnamefont {Iacovou}},
  \bibinfo {author} {\bibfnamefont {M.}~\bibnamefont {Suchak}}, \bibinfo
  {author} {\bibfnamefont {P.}~\bibnamefont {Bergstrom}}, \ and\ \bibinfo
  {author} {\bibfnamefont {J.}~\bibnamefont {Riedl}},\ }in\ \href {\doibase
  10.1145/192844.192905} {\emph {\bibinfo {booktitle} {Proceedings of the 1994
  ACM Conference on Computer Supported Cooperative Work}}},\ \bibinfo {series
  and number} {CSCW `94}\ (\bibinfo  {publisher} {ACM},\ \bibinfo {address}
  {New York (USA)},\ \bibinfo {year} {1994})\ pp.\ \bibinfo {pages}
  {175--186}\BibitemShut {NoStop}%
\bibitem [{Note5()}]{Note5}%
  \BibitemOpen
  \bibinfo {note} {\protect \url {http://movielens.org/}.}\BibitemShut {Stop}%
\bibitem [{Note6()}]{Note6}%
  \BibitemOpen
  \bibinfo {note} {``DeLorean'' by Aaron Humphreys; ``Darth Vader'' by Jake
  Dunham; ``Castle'' by Olly Banham; ``Movie Star'' by Nikita Kozin; ``Books on
  a Shelf'' by Lucas Glenn; ``Shark'' by Randomhero; ``Mask'' by Gorka Cestao;
  ``Zombie Hand'' by Valery; ``Army Helmet'' by Henry Ryder; ``Family'' by
  abeldb, from the Noun Project. All icons are under the CC
  licence.}\BibitemShut {Stop}%
\bibitem [{Note7()}]{Note7}%
  \BibitemOpen
  \bibinfo {note} {Every movie is assigned an array of 17 entries, representing
  the aforementioned genres. Each entry can be either zero or one, depending if
  that movie is considered as belonging to that genre or not (the number of
  ones in the vector can vary from 1 to a maximum of 6, if the selected film
  falls under several genres).}\BibitemShut {Stop}%
\bibitem [{\citenamefont {Hong}(2013)}]{Hong2013}%
  \BibitemOpen
  \bibfield  {author} {\bibinfo {author} {\bibfnamefont {Y.}~\bibnamefont
  {Hong}},\ }\href {\doibase 10.1016/j.csda.2012.10.006} {\bibfield  {journal}
  {\bibinfo  {journal} {Computational Statistics And Data Analysis}\ }\textbf
  {\bibinfo {volume} {59}},\ \bibinfo {pages} {41} (\bibinfo {year}
  {2013})}\BibitemShut {NoStop}%
\bibitem [{\citenamefont {Chen}\ \emph {et~al.}(1998)\citenamefont {Chen} \emph
  {et~al.}}]{Chen1998}%
  \BibitemOpen
  \bibfield  {author} {\bibinfo {author} {\bibfnamefont {S.~X.}\ \bibnamefont
  {Chen}} \emph {et~al.},\ }\href@noop {} {\bibfield  {journal} {\bibinfo
  {journal} {The Annals of Statistics}\ }\textbf {\bibinfo {volume} {26}},\
  \bibinfo {pages} {1894} (\bibinfo {year} {1998})}\BibitemShut {NoStop}%
\bibitem [{\citenamefont {Deheuvels}\ \emph {et~al.}(1989)\citenamefont
  {Deheuvels}, \citenamefont {Puri},\ and\ \citenamefont
  {Ralescu}}]{Deheuvels1989}%
  \BibitemOpen
  \bibfield  {author} {\bibinfo {author} {\bibfnamefont {P.}~\bibnamefont
  {Deheuvels}}, \bibinfo {author} {\bibfnamefont {M.~L.}\ \bibnamefont {Puri}},
  \ and\ \bibinfo {author} {\bibfnamefont {S.~S.}\ \bibnamefont {Ralescu}},\
  }\href {\doibase 10.1016/0047-259X(89)90111-5} {\bibfield  {journal}
  {\bibinfo  {journal} {Journal of Multivariate Analysis}\ }\textbf {\bibinfo
  {volume} {28}},\ \bibinfo {pages} {282} (\bibinfo {year} {1989})}\BibitemShut
  {NoStop}%
\bibitem [{\citenamefont {Volkova}(1996)}]{Volkova1996}%
  \BibitemOpen
  \bibfield  {author} {\bibinfo {author} {\bibfnamefont {A.~Y.}\ \bibnamefont
  {Volkova}},\ }\href {\doibase 10.1137/1140093} {\bibfield  {journal}
  {\bibinfo  {journal} {Theory of Probability And Its Applications}\ }\textbf
  {\bibinfo {volume} {40}},\ \bibinfo {pages} {791} (\bibinfo {year}
  {1996})}\BibitemShut {NoStop}%
\bibitem [{\citenamefont {Mikhailov}(1990)}]{Mikha1990}%
  \BibitemOpen
  \bibfield  {author} {\bibinfo {author} {\bibfnamefont {V.~G.}\ \bibnamefont
  {Mikhailov}},\ }\href@noop {} {\bibfield  {journal} {\bibinfo  {journal}
  {Theory of Probability and its Applications}\ }\textbf {\bibinfo {volume}
  {38}},\ \bibinfo {pages} {479} (\bibinfo {year} {1990})}\BibitemShut
  {NoStop}%
\end{thebibliography}%

\end{document}